\documentclass[journal]{IEEEtran}

\usepackage{graphicx}
\usepackage{color}
\usepackage{amsfonts}
\usepackage{float}
\usepackage{amsmath}
\usepackage{amsthm}
\usepackage{epstopdf}
\usepackage{changepage}
\usepackage{latexsym}
\usepackage{tikz}
\usepackage{algorithm,algorithmic}
\usepackage{amssymb}
\usepackage{multicol}
\usepackage{cases}
\usepackage{url}
\usepackage{xcolor}
\usepackage{booktabs}
\usepackage{multirow}
\usepackage{tgtermes}
\usepackage[utf8]{inputenc}
\usepackage[T1]{fontenc}
\usepackage{tabu}

\newcommand{\tabincell}[2]{\begin{tabular}{@{}#1@{}}#2\end{tabular}}

\floatstyle{plain}
\newfloat{myalgo}{tbhp}{mya}
{\begin{myalgo}[#1]
\centering
\begin{minipage}{#2}
\begin{algorithm}[H]}
{\end{algorithm}
\end{minipage}
\end{myalgo}}

\begin{document}

\title{Nonlinear Model Reduction in Power Systems by Balancing of Empirical Controllability and Observability Covariances}

\author{Junjian~Qi,~\IEEEmembership{Member,~IEEE,}
        Jianhui~Wang,~\IEEEmembership{Senior Member,~IEEE,}
        Hui Liu,~\IEEEmembership{Member,~IEEE, and}
        Aleksandar D. Dimitrovski,~\IEEEmembership{Senior Member,~IEEE}
        \thanks{This work was supported by the U.S. Department of Energy Office of Electricity Delivery and Energy Reliability. Paper no. TPWRS-00609-2015.

        J.~Qi and J. Wang are with the Energy Systems Division, Argonne National Laboratory, Argonne, IL 60439 USA (e-mails: jqi@anl.gov; jianhui.wang@anl.gov).
        
        H. Liu is with the Department of Electrical Engineering, Guangxi University, Nanning, 530004 China and was a visiting scholar at the Energy Systems Division, Argonne National Laboratory, Argonne, IL 60439 USA (e-mail: hughlh@126.com).
        
        A. D. Dimitrovski is with the Energy and Transportation Sciences Division, Oak Ridge National Laboratory, Oak Ridge, TN 37831 USA (e-mail: dimitrovskia@ornl.gov).
        
}
}

\markboth{preprint of doi: 10.1109/TPWRS.2016.2557760, IEEE Transactions on Power Systems.}{stuff}\maketitle

\begin{abstract}
In this paper, nonlinear model reduction for power systems is performed by the balancing of empirical controllability and observability covariances that are calculated around the operating region. 
Unlike existing model reduction methods, the external system does not need to be linearized but is directly dealt with as a nonlinear system. A transformation is found to balance the controllability and observability covariances in order to determine which states have the greatest contribution to the input-output behavior. The original system model is then reduced by Galerkin projection based on this transformation. The proposed method is tested and validated on a system comprised of a 16-machine 68-bus system and an IEEE 50-machine 145-bus system. 
The results show that by using the proposed model reduction the calculation efficiency can be greatly improved; at the same time, 
the obtained state trajectories are close to those for directly simulating the whole system or partitioning the system while not performing reduction. 
Compared with the balanced truncation method based on a linearized model, the proposed nonlinear model reduction method can guarantee higher accuracy and similar calculation efficiency. 
It is shown that the proposed method is not sensitive to the choice of the matrices for calculating the empirical covariances. 
\end{abstract}


\section{Introduction} \label{intro}

\IEEEPARstart{F}{aster} than real-time dynamic simulation can predict the dynamic system response to disturbances based on which 
the evaluation and analysis of outages including cascading blackouts \cite{us blackout}--\cite{bp16} can be performed and effective corrective actions can be identified \cite{sid}. 
However, large-scale power system dynamic simulation can involve several thousand state variables, and 
a detailed modeling of the whole system can lead to formidable computational burden. 
Dynamic model reduction, also known as dynamic equivalencing, is an effective approach for improving calculation efficiency and finally achieving faster than real-time simulation and control 
by reducing the external area to be a lower-order simpler model \cite{chow book}. 
Although the stability study by dynamic simulation is to determine the dynamic response of the generators and control systems in a study
area under disturbances inside the area, these disturbances will impact the neighboring area (called the external area), 
which in turn will impact the study area, due to the interconnected nature of large power systems. 

For model reduction, the study area is of interest and therefore is modeled in detail, while the external area 
is not of direct interest and thus can be reduced and replaced with a simpler mathematical description. 
Physically based coherency model reduction has been extensively studied \cite{chow book}--\cite{vittal1}; 
it first identifies coherency of generators and then performs reduction by aggregating the coherent generators. 
The performance of this method mainly depends on the identification of coherent generators. When system conditions change, 
it might be necessary to adjust the existing boundary to accurately capture the dynamic characteristics of the system \cite{vittal}, \cite{vittal1}.
Other approaches, such as synchrony \cite{syn}, singular perturbations \cite{singular}, selective modal analysis \cite{ma}, 
and computation intelligence methods \cite{ann1} have also been developed. 

There are also model reduction techniques based on the moment matching methods \cite{krylov1}--\cite{krylov}, which attempt to make the leading coefficients of a power series expansion of the reduced system's transfer function match those of the original system transfer function. 
Another model reduction approach from the perspective of input-output properties has also been studied, such as balanced truncation \cite{liu} 
and structured model reduction based on an extension balanced truncation \cite{luigi}. 
Compared with coherency-based methods, these methods have a stronger theoretical foundation and are more general, 
not specially targeted to a particular application \cite{luigi}. 

Besides, recently some new methods have also been developed, such as measurement-based model reduction \cite{meas1}--\cite{meas4}, 
border synchrony based method \cite{synchrony1}, ANN-based boundary matching technique \cite{ann}, independent component analysis approach \cite{independent}, 
heuristic optimization based approach \cite{heuristic1}, \cite{heuristic2}, and approximate bisimulation-based method \cite{bisimulation}. 
For detailed survey of the model reduction methods in power systems, the reader is referred to \cite{review1} and \cite{review2}. 

For most existing model reduction methods, the external system has to be linearized. 
Because of the strong nonlinearity of power systems, linearization-based methods cannot always provide accurate description of the physical system. 
In this paper, however, we discuss model reduction directly for nonlinear power systems through balanced truncation based on empirical controllability and observability covariances \cite{lall}--\cite{qi2}. This method has been discussed in \cite{lall}--\cite{hahn} where it has been applied to mechanical systems \cite{lall}, \cite{lall1} and chemical systems \cite{Gramianref}, \cite{hahn}. On one hand, similar to the balanced truncation method based on a linearized model, the proposed method also has a solid theoretical foundation and thus holds promise for 
application to large systems. On the other hand, the proposed method is expected to be able to perform more accurate model reduction by using the empirical controllability and observability covariances. Unlike analysis based on linearization, for which the controllability and observability only work locally in a neighborhood of an operating point, 
the empirical covariances are defined using the original system model and can thus reflect the controllability and observability of the full nonlinear dynamics in the given domain.

The remainder of this paper is organized as follows. 
Section \ref{covariance} introduces the empirical controllability and observability covariances and discusses their implementation. 
Section \ref{reduction} discusses the model reduction method based on the balancing of empirical controllability and observability covariances. 
Section \ref{red_power} applies the method in Section \ref{reduction} to the power system model. 
Section \ref{simu_whole} proposes a procedure for performing simulation for the study area and reduced external area. 
In Section \ref{case}, the proposed model reduction method is tested and validated on a system comprised of a 16-machine 68-bus system and an IEEE 50-machine 145-bus system. 
Finally, conclusions are drawn in Section \ref{conclusion}.

\section{Empirical Controllability and Observability Covariances} \label{covariance}

To perform model reduction for a system from the perspective of input-output properties, we should first obtain its input-output properties. 
For a linear time-invariant system
\begin{subnumcases} {\label{linear}}
\dot{\boldsymbol{x}}= \boldsymbol{A} \,\boldsymbol{x} + \boldsymbol{B} \,\boldsymbol{u} \\
\boldsymbol{y} = \boldsymbol{C} \,\boldsymbol{x} + \boldsymbol{D} \,\boldsymbol{u}
\end{subnumcases}
where $\boldsymbol{x} \in \mathbb{R}^n$ is the state vector, $\boldsymbol{u} \in \mathbb{R}^v$ is the input vector, 
and $\boldsymbol{y}\in \mathbb{R}^p$ is the output vector,
the controllability and observability gramians defined as \cite{kailath}
\begin{align}
\boldsymbol{W}_{\mathrm{c,L}}&=\int_0^\infty e^{\boldsymbol{A}\,t} \, \boldsymbol{B} \boldsymbol{B}^\top e^{\boldsymbol{A}^\top t}\,dt \\
\boldsymbol{W}_{\mathrm{o,L}}&=\int_0^\infty e^{\boldsymbol{A}^\top t}\,\boldsymbol{C}^\top \boldsymbol{C} \, e^{\boldsymbol{A}\,t}dt
\end{align}
can be used to analyze the controllability and observability and thus the input-state and state-output behavior. 
The gramians $\boldsymbol{W}_{\mathrm{c,L}}$ and $\boldsymbol{W}_{\mathrm{o,L}}$ are actually the unique positive definite solutions of the 
Lyapunov equations \cite{lall}
\begin{align}
\boldsymbol{A}\,\boldsymbol{W}_{\mathrm{c,L}}+\boldsymbol{W}_{\mathrm{c,L}} \, \boldsymbol{A}^\top + \boldsymbol{B}\,\boldsymbol{B}^\top &= 0 \\
\boldsymbol{A}^\top \,\boldsymbol{W}_{\mathrm{o,L}} + \boldsymbol{W}_{\mathrm{o,L}}\,\boldsymbol{A} + \boldsymbol{C}^\top \boldsymbol{C} &= 0.
\end{align} 

However, for a nonlinear system
\begin{subnumcases} {\label{n1}}
\dot{\boldsymbol{x}}=\boldsymbol{f}(\boldsymbol{x},\boldsymbol{u}) \\
\boldsymbol{y}=\boldsymbol{h}(\boldsymbol{x},\boldsymbol{u})
\end{subnumcases}
where $\boldsymbol{f}(\cdot)$ and $\boldsymbol{h}(\cdot)$ are the state transition and output functions, $\boldsymbol{x} \in \mathbb{R}^n$ is the state vector, 
$\boldsymbol{u} \in \mathbb{R}^v$ is the input vector, and $\boldsymbol{y}\in \mathbb{R}^p$ is the output vector, 
there is no analytical controllability or observability gramian.

In order to capture the controllability and observability of a nonlinear system, one can linearize the nonlinear system and calculate the gramians of the linearized system, 
in which case, however, the nonlinear dynamics of the system will be lost. 
Alternatively, in order to directly capture the input-output behavior of a nonlinear system in a similar way to a linear system, 
the empirical controllability and observability covariances \cite{lall}--\cite{qi2} are proposed, 
which provide a computable tool for empirical analysis of the input-state and state-output behavior of nonlinear systems, either by simulation or experiment.  

Different from analysis based on linearization, the empirical covariances are defined using the original system model 
and can thus reflects the controllability and observability of the full nonlinear dynamics in the given domain, 
whereas the controllability or observability gramians based on linearization only work locally in a neighborhood of an operating point. 
It is proven that the empirical covariances of a stable linear system described by (\ref{linear}) is equal to the usual gramians \cite{lall1}.

\subsection{Scaling the System} \label{scale}

The nonlinear system described by (\ref{n1}) should first be scaled because a state changing by orders of magnitude can be more important than a state that hardly changes, even though its steady state may have a smaller absolute value. Specifically, system (\ref{n1}) can be scaled by
\begin{align}
\tilde{\boldsymbol{x}}=\boldsymbol{T}_{x}^{-1}\,\boldsymbol{x}  \\
\tilde{\boldsymbol{u}}=\boldsymbol{T}_{u}^{-1}\,\boldsymbol{u} 
\end{align}
where $\boldsymbol{T}_{x}=\textrm{diag}(\boldsymbol{x}_{0})$, $\boldsymbol{T}_{u}=\textrm{diag}(\boldsymbol{u}_{0})$, $\boldsymbol{x}_{0}$ and 
$\boldsymbol{u}_{0}$ are the state and input at steady state, and the scaled system is
\begin{subnumcases} {\label{scale1}}
\dot{\tilde{\boldsymbol{x}}}=\boldsymbol{T}_{x}^{-1}\boldsymbol{f}(\boldsymbol{T}_{x}\,\tilde{\boldsymbol{x}},\boldsymbol{T}_{u}\,\tilde{\boldsymbol{u}}) \\
\boldsymbol{y}=\boldsymbol{h}(\boldsymbol{T}_{x}\,\tilde{\boldsymbol{x}},\boldsymbol{T}_{u}\,\tilde{\boldsymbol{u}}).
\end{subnumcases}

\subsection{Empirical Controllability Covariance}

The following sets are defined for empirical controllability covariance:
\begin{align}
&T^\mathrm{c}=\{\boldsymbol{T}_1^\mathrm{c},\cdots,\boldsymbol{T}_r^\mathrm{c};\,\boldsymbol{T}_l^\mathrm{c} \in \mathbb{R}^{v\times v},\,{\boldsymbol{T}_l^\mathrm{c}}^\top \boldsymbol{T}_l^\mathrm{c}=\boldsymbol{I}_v,\,l=1,\ldots,r\} \nonumber \\
&M^\mathrm{c}=\{c_1^\mathrm{c},\cdots,c_s^\mathrm{c};\,c_m^\mathrm{c} \in \mathbb{R},\,c_m^\mathrm{c}>0,\,m=1,\ldots,s\} \nonumber \\
&E^c=\{\boldsymbol{e}_1^\mathrm{c},\cdots,\boldsymbol{e}_v^\mathrm{c};\,\textrm{standard unit vectors in}\,\mathbb{R}^v\} \nonumber
\end{align}
where $r$ is the number of matrices for excitation directions, $s$ is the number of different excitation sizes for each direction,
and $v$ is the number of inputs to the system, and $\boldsymbol{I}_v$ is an identity matrix with dimension $v$.

For the nonlinear system described by (\ref{n1}), the empirical controllability covariance can be defined as 
\begin{equation} \label{cont}
\boldsymbol{W}_\mathrm{c}^{\mathrm{con}}=\sum_{i=1}^{v}\sum_{l=1}^r\sum_{m=1}^{s}\frac{1}{r\,s\,(c_m^c)^2}\int_0^\infty \boldsymbol{\Phi}^{ilm}(t)\, dt
\end{equation}
where $\boldsymbol{\Phi}^{ilm}(t)\in \mathbb{R}^{n\times n}$ is given by $\boldsymbol{\Phi}^{ilm}(t)=(\boldsymbol{x}^{ilm}(t)-\boldsymbol{x}_0^{ilm})(\boldsymbol{x}^{ilm}(t)-\boldsymbol{x}_0^{ilm})^\top$, 
$\boldsymbol{x}^{ilm}(t)$ is the state of the nonlinear system corresponding to the input $\boldsymbol{u}(t)=c_m^c \boldsymbol{T}_l^c \boldsymbol{e}_i \mathsf{v}(t)+\boldsymbol{u}_0(0)$, and $\mathsf{v}(t)$ is the shape of the input.

The discrete form of the empirical controllability covariance can be defined as \cite{Gramianref}
\begin{equation} \label{cont1}
\boldsymbol{W}_\mathrm{c}=\sum\limits_{i=1}^v\sum\limits_{l=1}^r\sum\limits_{m=1}^s \frac{1}{r\,s\,(c_m^c)^2}\sum\limits_{k=0}^K \boldsymbol{\Phi}_k^{ilm}\Delta t_k
\end{equation}
where $\boldsymbol{\Phi}_k^{ilm}\in \mathbb{R}^{n\times n}$ is given by $\boldsymbol{\Phi}_k^{ilm}=(\boldsymbol{x}_k^{ilm}-\boldsymbol{x}_0^{ilm})(\boldsymbol{x}_k^{ilm}-\boldsymbol{x}_0^{ilm})^\top$, 
$\boldsymbol{x}_k^{ilm}$ is the state of the nonlinear system at time step $k$ corresponding to the input $\boldsymbol{u}_k=c_m^c \boldsymbol{T}_l^c \boldsymbol{e}_i \mathsf{v}_k + \boldsymbol{u}_0(0)$, 
$K$ is the number of points chosen for the approximation of the integral in (\ref{cont}), and $\Delta t_k$ is the time interval between two points.

\subsection{Empirical Observability Covariance}

The following sets are defined for empirical observability covariances:
\begin{align}
T^o&=\{\boldsymbol{T}_1^o,\cdots,\boldsymbol{T}_r^o;\,\boldsymbol{T}_l^o \in \mathbb{R}^{n\times n},\,{\boldsymbol{T}_l^o}^\top \boldsymbol{T}_l^o=\boldsymbol{I}_n,\,l=1,\ldots,r\} \nonumber \\
M^o&=\{c_1^o,\cdots,c_s^o;\,c_m^o \in \mathbb{R},\;c_m^o>0,\,m=1,\ldots,s\} \nonumber \\
E^o&=\{\boldsymbol{e}_1^o,\cdots,\boldsymbol{e}_n^o;\,\textrm{standard unit vectors in}\,\mathbb{R}^n\} \nonumber
\end{align}
where $T^o$ defines the initial state perturbation directions, $r$ is the number of matrices for perturbation directions, $\boldsymbol{I}_n$ is an identity matrix with dimension $n$, 
$M^o$ defines the perturbation sizes and $s$ is the number of different perturbation sizes for each direction;
and $E^o$ defines the state to be perturbed and $n$ is the number of states of the system.

For the nonlinear system described by (\ref{n1}), the empirical observability covariance can be defined as
\begin{equation} \label{obse}
\boldsymbol{W}_o^{\mathrm{con}}=\sum_{l=1}^{r}\sum_{m=1}^{s}\frac{1}{r\,s\,(c_m^o)^2}\int_0^\infty \boldsymbol{T}_l^o \, \boldsymbol{\Psi}^{lm}(t) \, {\boldsymbol{T}_l^o}^\top dt
\end{equation}
where $\boldsymbol{\Psi}^{lm}(t)\in \mathbb{R}^{n\times n}$ is given by $\Psi_{ij}^{lm}(t)=(y^{ilm}(t)-y^{ilm,0})^\top (y^{jlm}(t)-y^{jlm,0})$, $y^{ilm}(t)$ is the output of the nonlinear system corresponding to the initial condition $\boldsymbol{x}(0)=c_m^o T_l^o e_i+\boldsymbol{x}_0$, and $y^{ilm,0}$ refers to the output measurement corresponding to the unperturbed initial state $\boldsymbol{x}_0$, which is usually chosen as the steady state under typical power flow conditions but can also be chosen as other operating points.

Similarly, (\ref{obse}) can be rewritten as its discrete form \cite{Gramianref}
\begin{equation} \label{obse1}
\boldsymbol{W}_o=\sum_{l=1}^{r}\sum_{m=1}^{s}\frac{1}{r\,s\,(c_m^o)^2}\sum_{k=0}^K \boldsymbol{T}_l^o \, \boldsymbol{\Psi}^{lm}_k \, {\boldsymbol{T}_l^o}^\top \Delta t_k
\end{equation}
where $\boldsymbol{\Psi}^{lm}_k \in \mathbb{R}^{n\times n}$ is given by ${\Psi^{lm}_k}_{ij}=(y^{ilm}_k-y^{ilm,0})^\top (y^{jlm}_k-y^{jlm,0})$, $y^{ilm}_k$ is the output at time step $k$, and $K$ and $\Delta t_k$ are the same as in (\ref{cont1}).

\section{Model Reduction by Balancing of Empirical Controllability and Observability Covariances} \label{reduction}

The empirical covariances obtained in Section \ref{covariance} contain important information about which states are controllable or observable, based on which a coordinate transformation 
$\boldsymbol{T} \in \mathbb{R}^{n\times n}$ can be obtained to transform the original model into another state space model whose states are decomposed into four categories: states which are 1) both controllable and observable; 2) controllable but not observable; 3) observable but not controllable; and 4) neither controllable nor observable. 

For the scaled system in (\ref{scale1}), let $\hat{\boldsymbol{x}}=\boldsymbol{T}\tilde{\boldsymbol{x}}$ and the transformed system is
\begin{subnumcases} {\label{trans}}
\dot{\hat{\boldsymbol{x}}}=\boldsymbol{T} \, \boldsymbol{T}_{x}^{-1} \boldsymbol{f}(\boldsymbol{T}_{x}\, \boldsymbol{T}^{-1}\, \hat{\boldsymbol{x}},\boldsymbol{T}_u \, \tilde{\boldsymbol{u}}) \\
\boldsymbol{y}=\boldsymbol{h}(\boldsymbol{T}_{x}\, \boldsymbol{T}^{-1}\, \hat{\boldsymbol{x}},\boldsymbol{T}_{u}\,\tilde{\boldsymbol{u}})
\end{subnumcases}
and the corresponding transformed covariances are
\begin{align} \label{trans cov1}
\boldsymbol{W}_c^{\textrm{tra}}&=\boldsymbol{T}\,\boldsymbol{W}_c\,\boldsymbol{T}^\top \\
\boldsymbol{W}_o^{\textrm{tra}}&=\big(\boldsymbol{T}^{-1}\big)^\top \,\boldsymbol{W}_o \,\boldsymbol{T}^{-1}.
\end{align} 

If the transformed covariances have the following feature
\begin{equation}
\boldsymbol{W}_c^{\textrm{tra}} = \left[ \begin{array}{cccc}
\boldsymbol{\Sigma}_1 & \boldsymbol{0} & \boldsymbol{0} & \boldsymbol{0} \\
\boldsymbol{0} & \boldsymbol{I} & \boldsymbol{0} & \boldsymbol{0} \\
\boldsymbol{0} & \boldsymbol{0} & \boldsymbol{0} & \boldsymbol{0} \\
\boldsymbol{0} & \boldsymbol{0} & \boldsymbol{0} & \boldsymbol{0}
\end{array} \right]
\end{equation}
\begin{equation}
\boldsymbol{W}_o^{\textrm{tra}} = \left[ \begin{array}{cccc}
\boldsymbol{\Sigma}_1 & \boldsymbol{0} & \boldsymbol{0} & \boldsymbol{0} \\
\boldsymbol{0} & \boldsymbol{0} & \boldsymbol{0} & \boldsymbol{0} \\
\boldsymbol{0} & \boldsymbol{0} & \boldsymbol{\Sigma}_3 & \boldsymbol{0} \\
\boldsymbol{0} & \boldsymbol{0} & \boldsymbol{0} & \boldsymbol{0}
\end{array} \right]
\end{equation}
where $\boldsymbol{\Sigma}_1$ and $\boldsymbol{\Sigma}_3$ are both diagonal matrices and $\boldsymbol{I}$ is an identity matrix, the transformed system in (\ref{trans}) is said to be \textit{balanced} and the corresponding transformed covariances are denoted by $\boldsymbol{W}_c^{\textrm{bal}}$ and $\boldsymbol{W}_o^{\textrm{bal}}$. 
The states of the balanced system are decoupled into the four categories mentioned above. 
Specifically, the covariance matrix of the states of the balanced system that are both controllable and 
observable is given by $\boldsymbol{\Sigma}_1$, the controllability covariance matrix of the states that are controllable but not observable is the identity matrix in the transformed controllability matrix, and the observability covariance matrix of the states that are observable but not controllable is $\boldsymbol{\Sigma}_3$ in the transformed observability matrix \cite{Gramianref}.

A proof for always existing a transformation that can balance a system is given in \cite{zhou}. 
As for how to calculate such a coordinate transformation $\boldsymbol{T}$ to balance a system that can be not completely controllable and observable, 
a method has been proposed in \cite{Gramianref}, which requires the calculation of four matrices 
$\boldsymbol{T}_1 \in \mathbb{R}^{n\times n}$, $\boldsymbol{T}_2 \in \mathbb{R}^{n\times n}$, $\boldsymbol{T}_3 \in \mathbb{R}^{n\times n}$, 
and $\boldsymbol{T}_4 \in \mathbb{R}^{n\times n}$ from the empirical covariances $\boldsymbol{W}_c$ and $\boldsymbol{W}_o$. 
In the following we will briefly introduce this method and more details can be found in \cite{Gramianref}. 

\begin{enumerate} \renewcommand{\labelitemi}{$\bullet$}
\vspace{0.2cm}
\item \textbf{Determine $\boldsymbol{T}_1$} 
\vspace{0.1cm}
\\
$\boldsymbol{T}_1$ is determined so that
\begin{equation}
\boldsymbol{T}_1\,\boldsymbol{W}_c \,\boldsymbol{T}_1^\top = \left[ \begin{array}{cc}
\boldsymbol{I}_c & \boldsymbol{0} \\
\boldsymbol{0} & \boldsymbol{0}
\end{array} \right]
\end{equation}
where $\boldsymbol{I}_c$ is an identity matrix with dimension equal to the rank of $\boldsymbol{W}_c$ and the rows and 
columns that contain only zeros refer to the rank deficiency of the controllability covariance.

\vspace{0.2cm}
\item \textbf{Determine $\boldsymbol{T}_2$} 
\vspace{0.1cm}
\\
The transformation $\boldsymbol{T}_1$ found in Step 1 is applied to the observability covariance
\begin{equation}
\boldsymbol{T}_1^\top \, \boldsymbol{W}_o \, \boldsymbol{T}_1^{-1} = \left[ \begin{array}{cc}
\tilde{\boldsymbol{W}}_{o,11} & \tilde{\boldsymbol{W}}_{o,12} \\
\tilde{\boldsymbol{W}}_{o,21} & \tilde{\boldsymbol{W}}_{o,22}
\end{array} \right]
\end{equation}
and a Schur decomposition can be found for the matrix $\tilde{\boldsymbol{W}}_{o,11}$ as
\begin{equation}
\boldsymbol{U}_1 \, \boldsymbol{W}_{o,11} \, \boldsymbol{U}_1^\top = \left[ \begin{array}{cc}
{\boldsymbol{\Sigma}_1}^2 & \boldsymbol{0} \\
\boldsymbol{0} & \boldsymbol{0}
\end{array} \right].
\end{equation}

The unitary matrix of this decomposition is required for the second part of the transformation and is given by
\begin{equation}
\big(\boldsymbol{T}_2^\top\big)^{-1} = \left[ \begin{array}{cc}
\boldsymbol{U}_1 & \boldsymbol{0} \\
\boldsymbol{0} & \boldsymbol{I}
\end{array} \right].
\end{equation}

\vspace{0.2cm}
\item \textbf{Determine $\boldsymbol{T}_3$} 
\vspace{0.1cm}
\\
A transformation using both $\boldsymbol{T}_1$ and $\boldsymbol{T}_2$ can be applied to 
the observability covariance matrix to obtain the third transformation, $\boldsymbol{T}_3$,
as given by
\begin{equation}
\big(\boldsymbol{T}_2^\top\big)^{-1}\,\big(\boldsymbol{T}_1^\top\big)^{-1}\,\boldsymbol{W}_o \,\boldsymbol{T}_1^{-1} \,\boldsymbol{T}_2^{-1} \notag
\end{equation}
\begin{equation} \qquad\quad\;\;\; = \left[ \begin{array}{ccc}
{\boldsymbol{\Sigma}_1}^2 & \boldsymbol{0} & \hat{\boldsymbol{W}}_{o,12} \\
\boldsymbol{0} & \boldsymbol{0} & \boldsymbol{0} \\
\hat{\boldsymbol{W}}_{o,12}^\top & \boldsymbol{0} & \hat{\boldsymbol{W}}_{o,22}
\end{array} \right]\qquad\qquad\;\;\;
\end{equation}
and 
\begin{equation}
\big(\boldsymbol{T}_3^\top\big)^{-1} = \left[ \begin{array}{ccc}
\boldsymbol{I} & \boldsymbol{0} & \boldsymbol{0} \\
\boldsymbol{0} & \boldsymbol{I} & \boldsymbol{0} \\
-\hat{\boldsymbol{W}}_{o,12}^\top \,{\boldsymbol{\Sigma}_1}^{-2} & \boldsymbol{0} & \boldsymbol{I}
\end{array} \right]. \quad
\end{equation}

\vspace{0.2cm}
\item \textbf{Determine $\boldsymbol{T}_4$} 
\vspace{0.1cm}
\\
A transformation using $\boldsymbol{T}_1$, $\boldsymbol{T}_2$, and $\boldsymbol{T}_3$ is applied to the observability covariance and a Schur decomposition is found for the square matrix containing the last columns and rows of the transformed system as
\begin{equation}
\hspace*{-0.2cm} \big(\boldsymbol{T}_3^\top\big)^{-1}\,\big(\boldsymbol{T}_2^\top\big)^{-1}\,\big(\boldsymbol{T}_1^\top\big)^{-1}\,\boldsymbol{W}_o \,\boldsymbol{T}_1^{-1} \,\boldsymbol{T}_2^{-1} \,\boldsymbol{T}_3^{-1} \notag
\end{equation}
\begin{equation} \hspace*{-0.09cm} = \left[ \begin{array}{ccc}
{\boldsymbol{\Sigma}_1}^2 & \boldsymbol{0} & \boldsymbol{0} \\
\boldsymbol{0} & \boldsymbol{0} & \boldsymbol{0} \\
\boldsymbol{0} & \boldsymbol{0} & \tilde{\boldsymbol{W}}_{o,22}-\hat{\boldsymbol{W}}_{o,12}^\top\,{\boldsymbol{\Sigma}_1}^{-2}\,\hat{\boldsymbol{W}}_{o,12}
\end{array} \right]\;
\end{equation}
and 
\begin{equation}
\boldsymbol{U}_2 \, \big(\tilde{\boldsymbol{W}}_{o,22}-\hat{\boldsymbol{W}}_{o,12}^\top \, {\boldsymbol{\Sigma}_1}^{-2} \, \hat{\boldsymbol{W}}_{o,12}\big) \, \boldsymbol{U}_2^\top \notag
\end{equation}
\begin{equation}
= \left[\begin{array}{cc}
\boldsymbol{\Sigma}_3 & \boldsymbol{0} \\
\boldsymbol{0} & \boldsymbol{0} \\
-\hat{\boldsymbol{W}}_{o,12}^\top \, {\boldsymbol{\Sigma}_1}^{-2} & \boldsymbol{0}
\end{array} \right]. \qquad\qquad\quad\;\;
\end{equation}

The forth transformation can further be determined by 
\begin{equation}
\big(\boldsymbol{T}_4^\top\big)^{-1} = \left[ \begin{array}{ccc}
{\boldsymbol{\Sigma}_1}^{-1/2} & \boldsymbol{0} & \boldsymbol{0} \\
\boldsymbol{0} & \boldsymbol{I} & \boldsymbol{0} \\
\boldsymbol{0} & \boldsymbol{0} & \boldsymbol{U}_2
\end{array} \right]. \qquad
\end{equation}

\end{enumerate}

Then the transformation matrix $\boldsymbol{T}$ that balances the states that are observable and controllable is given by
\begin{equation} \label{Tbal}
\boldsymbol{T}=\boldsymbol{T}_4\,\boldsymbol{T}_3\, \boldsymbol{T}_2\, \boldsymbol{T}_1
\end{equation}
which can be further used to reduce the scaled system in (\ref{scale1}) by Galerkin projection \cite{Gramianref}, \cite{hahn}.
Specifically, let $\bar{\boldsymbol{x}}=\boldsymbol{T} \, \tilde{\boldsymbol{x}}$ and the reduced system is
\begin{subnumcases} {\label{reduce}}
\dot{\bar{\boldsymbol{x}}}_1 = \boldsymbol{P} \, \boldsymbol{T} \,\boldsymbol{T}_{x}^{-1} \boldsymbol{f}(\boldsymbol{T}_{x} \, \boldsymbol{T}^{-1} \, \bar{\boldsymbol{x}},\boldsymbol{T}_u\,\tilde{\boldsymbol{u}}) \\
\bar{\boldsymbol{x}}_2= \bar{\boldsymbol{x}}_{2ss} \\
\boldsymbol{y}=\boldsymbol{h}(\boldsymbol{T}_{x} \, \boldsymbol{T}^{-1} \, \bar{\boldsymbol{x}},\boldsymbol{T}_{u} \, \tilde{\boldsymbol{u}})
\end{subnumcases}
where $\boldsymbol{P}=[\boldsymbol{I}_{n_{\textrm{red}}} \; \boldsymbol{0}]$ is the projection matrix, which has the rank of the reduced system $n_{\textrm{red}}$; 
$\bar{\boldsymbol{x}}_1$ and $\bar{\boldsymbol{x}}_2$ respectively represent the retained states and the reduced states, 
among which $\bar{\boldsymbol{x}}_2$ are kept at their steady state values $\bar{\boldsymbol{x}}_{2ss}$. 

Here, $n_{\textrm{red}}$ can be determined by Hankel singular values, which are the eigenvalues of $\boldsymbol{W}_o^{\textrm{bal}} \boldsymbol{W}_c^{\textrm{bal}}$ \cite{lall}--\cite{hahn}. 
The Hankel singular values provide a measure for the importance of the states in the sense that the state with the largest singular value is affected the most by the control inputs and the output is most affected by the change of this state. 
Thus the states corresponding to the largest singular values influence the input-output behavior the most. 
When the states that correspond to zero or very small Hankel singular values are eliminated, 
the reduced system retains most of the input-output behavior of the full-order system.

\section{Reduction for Power System Model} \label{red_power}

\begin{figure}[!t]
\centering
\includegraphics[width=3.3in]{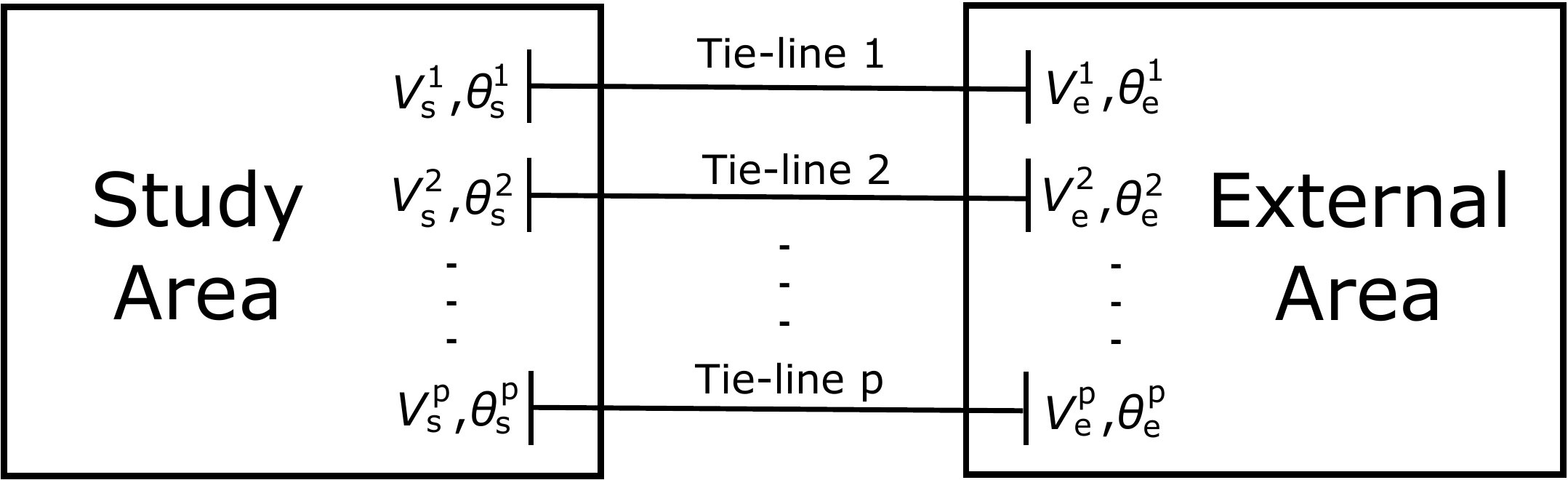}
\caption{System configuration of the study area and external area.}
\label{config}
\end{figure}

The whole system is partitioned into the study area and external area (see Fig. \ref{config}). 
The study area has $n_g^s$ generators and $n_b^s$ buses and the external area has $n_g^e$ generators and $n_b^e$ buses. 
There are $p$ tie-lines between the study and external area, and the set of boundary buses that belong to the study and external area are denoted by $\mathcal{B}_{s,\textrm{bound}}=\{b_1^s,b_2^s,\cdots,b_s^p$\} and $\mathcal{B}_{e,\textrm{bound}}=\{b_1^e,b_2^e,\cdots,b_p^e\}$. Correspondingly, the voltage magnitude and phase angles of the boundary bus $b_i^s,i\in \{1,2,\cdots,p\}$ are denoted by $V_i^s$ and $\theta_i^s$, 
and those for the boundary bus $b_i^e,i\in \{1,2,\cdots,p\}$ are denoted by $V_i^e$ and $\theta_i^e$.

The model reduction method in Section \ref{reduction} is applied to reduce the external area. The model reduction procedure can be summarized in the following four steps.

\begin{enumerate} \renewcommand{\labelitemi}{$\bullet$}
\vspace{0.2cm}
\item \textbf{Scale the external system} 
\vspace{0.1cm}
\\
The external system is scaled by using the method in Section \ref{scale}. 

\vspace{0.2cm}
\item \textbf{Calculate empirical covariances} 
\vspace{0.1cm}
\\
The empirical controllability and observability covariances are calculated for the scaled system on time interval $[0,t_f]$.
In (\ref{cont1}) and (\ref{obse1}) $\Delta t_k$ can take different values according to the required accuracy, and 
$\boldsymbol{x}_0$ is the steady state. 

For the external area, the inputs and outputs are, respectively, the voltage magnitude and the phase angles of the boundary buses in $\mathcal{B}_{s,\textrm{bound}}$ and $\mathcal{B}_{e,\textrm{bound}}$. 
More details about the power system model can be found in Appendices \ref{appA} and \ref{appB}.

\vspace{0.2cm}
\item \textbf{Balance empirical covariances} 
\vspace{0.1cm}
\\
The balancing of empirical covariances is performed as discussed in Section \ref{reduction} and 
the coordinate transformation that can balance the scaled external system is obtained by (\ref{Tbal}).

\vspace{0.2cm}
\item \textbf{Perform model reduction} 
\vspace{0.1cm}
\\
Model reduction is performed for the external area by (\ref{reduce}).

\end{enumerate}

\section{Simulation of the Whole System} \label{simu_whole}

The whole system is partitioned into the study area and the external area, as shown in Fig. \ref{config}.
For both areas, the boundary buses in the other area are treated as generators with a classical second-order model and very large inertia constant.
The generators corresponding to boundary buses that belong to the study area and external area are denoted by sets 
$\mathcal{G}_s=\{g_1^s,g_2^s,\cdots,g_p^s$\} and $\mathcal{G}_e=\{g_1^e,g_2^e,\cdots,g_p^e\}$. The whole system can be simulated in the following way.

\begin{enumerate} \renewcommand{\labelitemi}{$\bullet$}
\vspace{0.2cm}
\item \textbf{Simulate the study area} 
\vspace{0.1cm}
\\
The simulation is performed for the study area, the tie-lines, and the boundary buses in the external area.
Since the boundary buses $b_1^e,b_2^e,\cdots,b_p^e$ are treated as generators, the simulated system thus has a total of $n_g^s+p$ generators and $n_b^s+p$ buses.

The states of the study area at time step $k+1$, denoted by $\boldsymbol{x}_{s,k+1}$, can be obtained by solving the following differential equations
\begin{equation} \label{study_solve}
\dot{\boldsymbol{x}}_s=\boldsymbol{f}_s(\boldsymbol{x}_s,\boldsymbol{u}_s)
\end{equation}
with given $\boldsymbol{x}_{s,k}$ that is the state at time step $k$.

The input $\boldsymbol{u}_s$ is comprised of voltage magnitude and phase angles of the boundary buses in $\mathcal{B}_{e,\textrm{bound}}$ and 
can be written as $\boldsymbol{u}_{s,k}=\big[\boldsymbol{V}_{e,k}^\top \;\; \boldsymbol{\theta}_{e,k}^\top\big]^\top$ for time step $k$.

When solving (\ref{study_solve}), since only the second-order generator model is used, the voltage magnitude of the boundary buses (also transient voltage $e'_\mathrm{q}$ of the corresponding generators) will remain unchanged. In addition, since the inertia constant is very large, the phase angle of the boundary buses (also rotor angle $\delta$ of the corresponding generators) will not change.

The rotor angle and transient voltage at $q$ and $d$ axes at time step $k+1$ of the generators in study area (not including boundary buses in external area) are denoted by $\boldsymbol{\delta}_{s,k+1}$, $\boldsymbol{e'_\mathrm{q}}_{s,k+1}$, and $\boldsymbol{e'_\mathrm{d}}_{s,k+1}$.

\vspace{0.2cm}
\item \textbf{Simulate the external area} 
\vspace{0.1cm}
\\
The simulation is performed for the external area, the tie-lines, and the boundary buses in the study area.
The boundary buses $b_1^s,b_2^s,\cdots,b_p^s$ are treated in the same way as in Step 1 and
the simulated system thus has a total of $n_g^e+p$ generators and $n_b^e+p$ buses.

The states of the reduced external system at time step $k+1$, denoted by $\bar{\boldsymbol{x}}_{e1,k+1}$, can be obtained by
solving the differential equations
\begin{equation} \label{exter_solve}
\dot{\bar{\boldsymbol{x}}}_{e1}=\boldsymbol{P} \, \boldsymbol{T} \, \boldsymbol{T}_{x}^{-1} \boldsymbol{f}_e(\boldsymbol{T}_{x} \, \boldsymbol{T}^{-1} \bar{\boldsymbol{x}}_e,\boldsymbol{u}_e)
\end{equation}
with given $\bar{\boldsymbol{x}}_{e,k}$, state of external area at time step $k$.

The input $\boldsymbol{u}_e$ is comprised of voltage magnitude and phase angles of the boundary buses in $\mathcal{B}_{s,\textrm{bound}}$ and 
can be written as $\boldsymbol{u}_{e,k}=\big[\boldsymbol{V}_{s,k}^\top \;\; \boldsymbol{\theta}_{s,k}^\top\big]^\top$ for time step $k$.
Similar to Step 1, the voltage magnitude and phase angles of the boundary buses will remain unchanged.

The states of the original system can be obtained by transformation of the states of the reduced external system as $\boldsymbol{x}_e=\boldsymbol{T}_x \, \boldsymbol{T}^{-1}\big[\bar{\boldsymbol{x}}_{e1}^\top \;\; \bar{\boldsymbol{x}}_{e2ss}^\top\big]^\top$. 
The rotor angle at time step $k+1$ of the generators in external area (not including boundary buses in study area)
is denoted by $\boldsymbol{\delta}_{e,k+1}$. The transient voltages at $q$ and $d$ axes are denoted by $\boldsymbol{e'_q}_{e,k+1}$ and $\boldsymbol{e'_d}_{e,k+1}$.

\vspace{0.2cm}
\item \textbf{Update boundary buses} 
\vspace{0.1cm}
\\
Given the states of the study area $\boldsymbol{\delta}_{s,k+1}$, $\boldsymbol{e'_\mathrm{q}}_{s,k+1}$, and $\boldsymbol{e'_\mathrm{d}}_{s,k+1}$ and the states of the external area $\boldsymbol{\delta}_{e,k+1}$ at time step $k+1$, the voltage sources of the generators can be obtained as follows:
\begin{subequations}
\begin{align}
\hspace*{-0.2cm}&\boldsymbol{\it \Psi}_e^{re} = \boldsymbol{e'_\mathrm{d}}_{e,k+1} \sin\boldsymbol{\delta}_{e,k+1}  + \boldsymbol{e'_\mathrm{q}}_{e,k+1} \cos\boldsymbol{\delta}_{e,k+1}  \\
&\boldsymbol{\it \Psi}_e^{im} =  \boldsymbol{e'_\mathrm{q}}_{e,k+1} \sin\boldsymbol{\delta}_{e,k+1} - \boldsymbol{e'_\mathrm{d}}_{e,k+1} \cos\boldsymbol{\delta}_{e,k+1}  \\
&\boldsymbol{\it \Psi}_e^{\textrm{state}} = \boldsymbol{\it \Psi}_e^{re} + j \boldsymbol{\it \Psi}_e^{im} \\
&\boldsymbol{\it \Psi}_e^{\textrm{input}} = \boldsymbol{V}_{s,k+1}\, e^{\,j \boldsymbol{\theta}_{s,k+1}} \\
&\boldsymbol{\it \Psi}_s^{re} = \boldsymbol{e'_\mathrm{d}}_{s,k+1} \sin\boldsymbol{\delta}_{s,k+1} + \boldsymbol{e'_\mathrm{q}}_{s,k+1} \cos \boldsymbol{\delta}_{s,k+1}  \\
&\boldsymbol{\it \Psi}_s^{im} =  \boldsymbol{e'_\mathrm{q}}_{s,k+1} \sin\boldsymbol{\delta}_{s,k+1} - \boldsymbol{e'_\mathrm{d}}_{s,k+1} \cos \boldsymbol{\delta}_{s,k+1} \\
&\boldsymbol{\it \Psi}_s^{\textrm{state}} = \boldsymbol{\it \Psi}_s^{re} + j \boldsymbol{\it \Psi}_s^{im} \\
&\boldsymbol{\it \Psi}_s^{\textrm{input}} = \boldsymbol{V}_{e,k+1}\, e^{\,j \boldsymbol{\theta}_{e,k+1}}.
\end{align}
\end{subequations}

As in Appendix \ref{appA}, we denote by $\mathcal{B}_{s,\textrm{ZIP}}$ the $n_{\textrm{ZIP}}^s$ load buses in the study area that are modeled as ZIP load (also called non-conforming load, as in \cite{pst}). The other buses are denoted by $\mathcal{B}_{s,\textrm{ZIP}}^c$ and all of the buses are $\mathcal{B}_s$.

The voltage reconstruction matrix for the study area (including the boundary buses in the other area),  which gives the original bus voltages components due to the generator internal bus voltages,
is denoted by $\boldsymbol{R}_{gs} \in \mathbb{C}^{(n_b^s+p-n_{\textrm{ZIP}}^s)\times (n_g^s+p)}$.
\begin{align}
&\tilde{\boldsymbol{V}}_{s,\mathcal{B}_{s,\textrm{ZIP}}}=\tilde{\boldsymbol{V}}_{ncs} \label{vnc1} \\
&\tilde{\boldsymbol{V}}_{s,\mathcal{B}_{s,\textrm{ZIP}}^c}=\boldsymbol{R}_{gs} \big[{\boldsymbol{\it \Psi}_s^{\textrm{state}}}^\top \; {\boldsymbol{\it \Psi}_s^{\textrm{input}}}^\top\big]^\top + \boldsymbol{R}_{ncs} \tilde{\boldsymbol{V}}_{ncs} \label{vnc2}
\end{align}
where $\tilde{\boldsymbol{V}}_s$ is the complex voltages for all buses in $\mathcal{B}_s$, $\tilde{\boldsymbol{V}}_{s,\mathcal{B}_{s,\textrm{ZIP}}}$ and $\tilde{\boldsymbol{V}}_{s,\mathcal{B}_{s,\textrm{ZIP}}^c}$ are, respectively, the complex voltages for the non-conforming load buses and the other buses, $\boldsymbol{R}_{ncs} \in \mathbb{C}^{(n_b^s+p-n_{\textrm{ZIP}}^s)\times n_{\textrm{ZIP}}^s}$ is the voltage reconstruction matrix which gives the original bus voltages components due to the non-conforming load, and $\tilde{\boldsymbol{V}}_{ncs} \in \mathbb{C}^{n_{\textrm{ZIP}}^s \times 1}$ is the complex voltages of the non-conforming load buses that can be obtained as $\tilde{\boldsymbol{V}}_{nc}$ by solving the nonlinear equations in (\ref{vnc}) by Newton's method. 
Similarly, we can also get $\tilde{\boldsymbol{V}}_{e,\mathcal{B}_{e,\textrm{ZIP}}}$ and $\tilde{\boldsymbol{V}}_{e,\mathcal{B}_{e,\textrm{ZIP}}^c}$ for the external area for which the notations are similar to those for the study area.

Then the nonlinear equations for the boundary buses at time step $k+1$ can be written as follows, for which $\boldsymbol{V}_{s,k+1}$, $\boldsymbol{V}_{e,k+1}$, $\boldsymbol{\theta}_{s,k+1}$, and $\boldsymbol{\theta}_{e,k+1}$ are unknowns:
\begin{equation}
\Biggr| \left[ \begin{array}{c}
\tilde{\boldsymbol{V}}_{s,\mathcal{B}_{s,\textrm{bound}}} \\
\tilde{\boldsymbol{V}}_{e,\mathcal{B}_{e,\textrm{bound}}}
\end{array} \right] \Biggr| =\left[ \begin{array}{c}
\boldsymbol{V}_{s,k+1} \\
\boldsymbol{V}_{e,k+1}
\end{array} \right]
\end{equation}
\begin{equation}
\arg\Biggr(\left[ \begin{array}{c}
\tilde{\boldsymbol{V}}_{s,\mathcal{B}_{s,\textrm{bound}}} \\
\tilde{\boldsymbol{V}}_{e,\mathcal{B}_{e,\textrm{bound}}}
\end{array} \right]\Biggr)=\left[ \begin{array}{c}
\boldsymbol{\theta}_{s,k+1} \\
\boldsymbol{\theta}_{e,k+1}
\end{array} \right]
\end{equation}
where $\tilde{\boldsymbol{V}}_{s,\mathcal{B}_{s,\textrm{bound}}}$ and $\tilde{\boldsymbol{V}}_{e,\mathcal{B}_{e,\textrm{bound}}}$ are, respectively, the complex voltages of the boundary buses in the study area and external area that are obtained by (\ref{vnc1})-(\ref{vnc2}), and $|\cdot|$ and $\arg(\cdot)$ represent the absolute value and argument of a complex vector. Note that the left-hand side of these equations are actually also functions of the unknowns $\boldsymbol{V}_{s,k+1}$, $\boldsymbol{V}_{e,k+1}$, $\boldsymbol{\theta}_{s,k+1}$, and $\boldsymbol{\theta}_{e,k+1}$.

The obtained nonlinear equations can be solved by Newton's method, for which the inputs $\boldsymbol{u}_{s,k}$ and $\boldsymbol{u}_{e,k}$ at time step $k$ are used as initial guess. The solution of the nonlinear equations can be used to update $\boldsymbol{u}_{s,k+1}$ and $\boldsymbol{u}_{e,k+1}$, which are further used for simulation in Steps 1 and 2 for the next time step.

\end{enumerate}

\section{Case Studies} \label{case}

The proposed model reduction method is tested on a system comprised of a 16-machine 68-bus system as the study area 
and an IEEE 50-machine 145-bus system as the external area. 
Both systems are extracted from Power System Toolbox \cite{pst}. 
The empirical covariance calculation and model reduction are implemented with Matlab.
All tests are carried out on a 3.2-GHz Intel(R) Core(TM) i7-4790S based desktop. 

For the study area, the fast sub-transient dynamics and saturation effects are ignored and the generators are described by the two-axis transient model with IEEE Type DC1 excitation system. Each generator has seven state variables, which are rotor angle $\delta$, rotor speed $\omega$, transient voltage along $\mathrm{q}$ and $\mathrm{d}$ axes $e'_{\mathrm{q}i}$ and $e'_{\mathrm{d}i}$, regulator output voltage $V_\mathrm{R}$, excitation output voltage $E_{\mathrm{fd}}$, and stabilizing transformer state variable $R_\mathrm{f}$. A subset of load buses, buses 1, 16, 23, 28, 39, 45, 48, and 51, are modeled as ZIP loads. 
The proportions of constant impedance, constant current, and constant power loads are determined by the parameters $p_1$, $p_2$, $p_3$, $q_1$, $q_2$, and $q_3$ in Appendix \ref{appA}. 
We choose $p_1=q_1=0.2$, $p_2=q_2=0.3$, and $p_3=q_3=0.5$. The other loads are modeled as constant impedance. More load buses can be modeled as ZIP loads. But there is a tradeoff between the model accuracy and the computational complexity, since the computation burden of both the differential equations and the boundary bus updating will increase when the number of ZIP loads increases.

For the external system extracted from PST, only seven generators (generators 1--6 and 23)
have high-order model while all the others only use a second-order model. 
Here, we use a fourth-order transient model to describe generators 1--6 and 23, for which the state variables are rotor angle $\delta$ and rotor speed $\omega$, 
and transient voltage along $\mathrm{q}$ and $\mathrm{d}$ axes $e'_{\mathrm{q}}$ and $e'_{\mathrm{d}}$, and a second-order classical model for the others, for which the state variables are rotor angle $\delta$ and rotor speed $\omega$. All of the loads are modeled as constant impedance.
More details about the models for the study and external areas can be found in Appendices \ref{appA} and \ref{appB}.

\subsection{Parameter Setup} \label{para_set}

The $\Delta t_k$ in (\ref{cont1}) and (\ref{obse1}) is chosen as $0.01$s.
The empirical controllability and observability covariances are calculated for the scaled system in time interval $[0,5\,\textrm{s}]$. 
When calculating empirical controllability or observability covariance, the inputs or the states are perturbed by adding a step change at $t=0$.  
For $T^c$ and $T^o$, a reasonably simple choice is 
\begin{align}
T^c &=\{\boldsymbol{I}_v,-\boldsymbol{I}_v\} \\
T^o &=\{\boldsymbol{I}_n,-\boldsymbol{I}_n\}
\end{align}
where $\boldsymbol{I}_v$ and $\boldsymbol{I}_n$ are identity matrix with dimension $v$ and $n$, 
since this corresponds to using both positive and negative inputs or initial states perturbations 
on each input or each state separately \cite{lall}. 
For $M^c$ and $M^o$, we first choose a linearly scaled set $M_0=\{0.25, 0.5, 0.75, 1.0\}$ and let 
\begin{align}
&M^c_{u}=k_{u} M_0 \label{Mc} \\
&M^o_{x}=k_{x} M_0 \label{Mo}
\end{align}
where $u$ is an input of the external area and can be $V$ or $\theta$, 
$x$ is a state variable of the external area that can be 
$\delta$, $\omega$, $e'_\mathrm{q}$, or $e'_\mathrm{d}$, and $k_{u}$ and $k_{x}$ are used to 
consider different ranges of change for different types of variables. 
For example, the voltage magnitude can only change in a small range while phase angle can change much more significantly. 
Then the perturbation for $u$ or $x$ will range from 25$k_{u}$\% or 25$k_{x}$\% to 100$k_{u}$\% or 100$k_{x}$\% of the steady state value. 

In order to determine $k_{u}$ and $k_{x}$, we apply a total of $n_\mathrm{f}=100$ three-phase faults, for each of which the fault is applied on one of the randomly chosen lines at one end and 
is cleared at near and remote end after $0.05$s and $0.1$s. 
For a fault $j$, we calculate the changes from the pre-fault input $u_{\mathrm{e}i0}$ or state $x_{\mathrm{e}i0}$ to 
the post-fault input $u_{\mathrm{e}i\mathrm{f}}$ or state $x_{\mathrm{e}i\mathrm{f}}$ for the $i$th input or state as 
\begin{align}
\Delta u_{\mathrm{e}i}^j &= \frac{u_{\mathrm{e}i\mathrm{f}}-u_{ei0}}{u_{\mathrm{e}i\mathrm{0}}} \\
\Delta x_{\mathrm{e}i}^j &= \frac{x_{\mathrm{e}i\mathrm{f}}-x_{\mathrm{e}i0}}{x_{\mathrm{e}i0}}. 
\end{align}
The $k_u$ and $k_x$ can thus be calculated as
\begin{align}
k_u &= \alpha_u \cdot \frac{1}{p}\sum\limits_{i=1}^{p} ||\Delta \boldsymbol{u}_{\mathrm{e}i}||_\infty \\
k_x &= \alpha_x \cdot \frac{1}{n_x}\sum\limits_{i=1}^{n_x} ||\Delta \boldsymbol{x}_{\mathrm{e}i}||_\infty 
\end{align}
where $p$ is the number of inputs of the external area, $n_x$ is the number of generators with state variable $x$ in the external area, 
$\Delta \boldsymbol{u}_{\mathrm{e}i} = \big[\Delta u_{\mathrm{e}i}^1, \cdots, \Delta u_{\mathrm{e}i}^{n_\mathrm{f}}\big]^\top$, $\Delta \boldsymbol{x}_{\mathrm{e}i} = \big[\Delta x_{\mathrm{e}i}^1, \cdots, \Delta x_{ei}^{n_\mathrm{f}}\big]^\top$,
$||\boldsymbol{v}||_\infty$ is the infinity norm of a $n$-dimensional vector $\boldsymbol{v}$ defined as 
\begin{equation}
||\boldsymbol{v}||_\infty=\max\big(|v_1|,\cdots,|v_n|\big),
\end{equation}
and $\alpha_u$ and $\alpha_x$ are chosen as real numbers greater than 1.0 (here we choose them as 2) since the applied $n_f$ faults cannot represent all of the possible disturbances.
By using this method, $k_{u}$ and $k_{x}$ are determined, as listed in Table \ref{k}, which shows that 
different types of variables do have very different ranges of change.

\begin{table}[!t]
\renewcommand{\arraystretch}{1.4}
\caption{The Determined $k_{u}$ and $k_{x}$}
\label{k}
\centering 
\begin{tabu}{cccccc}
\hline
     $k_{V}$  & $k_{\theta}$  & $k_{\delta}$ & $k_{\omega}$ & $k_{e'_q}$ & $k_{e'_d}$  \\
    \midrule
    0.054 & 1.24 & 0.90 & 0.0050 & 0.024 & 0.27  \\
\hline
\end{tabu}
\end{table}

\subsection{Scenario Setup}

\begin{figure}[!t]
	\centering
	\includegraphics[width=3.5in]{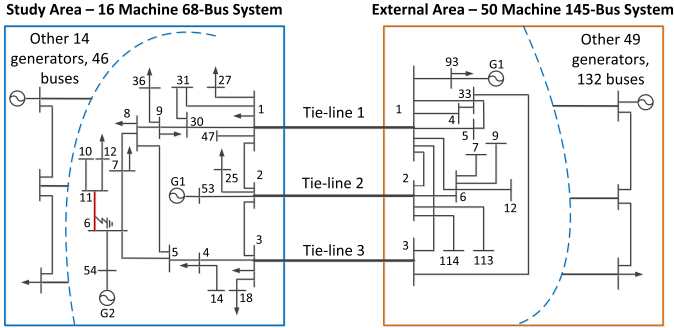}
	\caption{Test system with three tie-lines. The study area is 16-machine 68-bus system and the external area is the IEEE 50-machine 145-bus system. 
	The location where a three-phase fault is applied is highlighted by red line. }
	\label{system}
\end{figure}

Without losing generality, we add three tie-lines between the study and the external area which connect bus $i$ in study area to bus $i$ in external area, where $i=1,2,3$. 
To generate dynamic response, a three-phase fault is applied at bus $6$ of line $6-11$ in the study area at $0.1$s and is cleared at the near and remote ends after $0.05$s and $0.1$s. 
The corresponding test system and the location where the fault is applied are shown in Fig. \ref{system}. 
For simplicity, we only show the parts of the study area and the external area that are close to the boundary buses. 
The simulation is performed for $15$ seconds and the time step is $0.01$s and $0.03$s, respectively, for before and after the fault clearing. 
The differential equations are solved by Matlab function ``$\operatorname{ode23t}$''. 

Note that the dynamic simulation is performed for $15$ seconds while the empirical controllability and observability covariance calculation is only for the first $5$ seconds. 
In the following sections we will show that the empirical covariances obtained in this manner are good enough for performing model reduction for the external area.

It has been shown in  \cite{chow book} that the reduced-order model via balanced truncation \cite{liu} represents a better approximation with lower orders compared with the Krylov subspace method \cite{krylov}. Thus we only compare the proposed method with the balanced truncation method using a linearized model in \cite{liu}. 

The external area has $G_e=50$ generators. Seven of them have fourth-order transient model and the others have second-order classical model. 
Therefore, there are a total of 114 state variables. 
The number of retained states $n_{\textrm{red}}$ can be determined by Hankel singular values. 
For our test case, only 9 of the Hankel singular values are greater than $10^{-5}$ and we thus choose $n_{\textrm{red}}=9$, 
which only accounts for 7.9\% of the number of states and is also used for the method in \cite{liu}. 

Note that we apply the method in Section \ref{reduction} to calculate the transformation matrix $\boldsymbol{T}$ for the balanced truncation method based on a linearized model in \cite{liu}, 
rather than directly using the method used in \cite{liu}, which is proposed in \cite{laub} and can be summarized as:
\begin{align}
&\boldsymbol{W}_c=\boldsymbol{L}_c \boldsymbol{L}_c^\top \\
&\boldsymbol{W}_o=\boldsymbol{L}_o \boldsymbol{L}_o^\top \\
& \boldsymbol{L}_o^\top \boldsymbol{L}_c = \boldsymbol{U} \boldsymbol{\Lambda} \boldsymbol{V}^\top \\
& \boldsymbol{T}=\boldsymbol{L}_c \boldsymbol{V} \boldsymbol{\Lambda}^{-1/2}.
\end{align}
If the transformation matrix obtained by this method is used to get the reduced model for the linearized system, 
the corresponding simulation using the reduced model cannot proceed because the Newton's method is difficult to converge when used to solve the nonlinear equations in (\ref{vnc}). 
By contrast, by using the method in Section \ref{reduction} to get the transformation matrix and further getting the reduced model of the linearized system, 
the performance of the simulation is acceptable, although not as good as that of the proposed nonlinear model reduction method. 
This is mainly because the balancing transformation method discussed in Section \ref{reduction} is applicable to systems that are not completely controllable and observable \cite{Gramianref}.

The simulation methods considered in this paper are summarized in Table \ref{scenario}. The results for these methods will be given in the following sections.

\begin{table}[!t]
\renewcommand{\arraystretch}{1.5}
\caption{Simulation Methods}
\label{scenario}
\centering
\begin{tabular}{cc}
\hline
Method & Definition \\
\hline
UnPartitioned & \tabincell{l}{Simulate the whole system without partition} \\
\hline
Partitioned-Unreduced & \tabincell{l}{Partition the whole system into \\ study area and external area, \\ while not reducing the external system}    \\
\hline
Partitioned-Reduced-NM & \tabincell{l}{Partition the whole system and reduce \\ the external area by the proposed method \\  based on the \textbf{N}onlinear \textbf{M}odel (NM)}  \\
\hline
Partitioned-Reduced-LM & \tabincell{l}{Partition the whole system and reduce \\  the external area by method in \cite{liu} \\  based on the \textbf{L}inearized \textbf{M}odel (LM)}   \\
\hline
\end{tabular}
\end{table}

\subsection{Results for the Study Area} \label{study_result}

\begin{figure}[!t]
\centering
\includegraphics[width=3.4in]{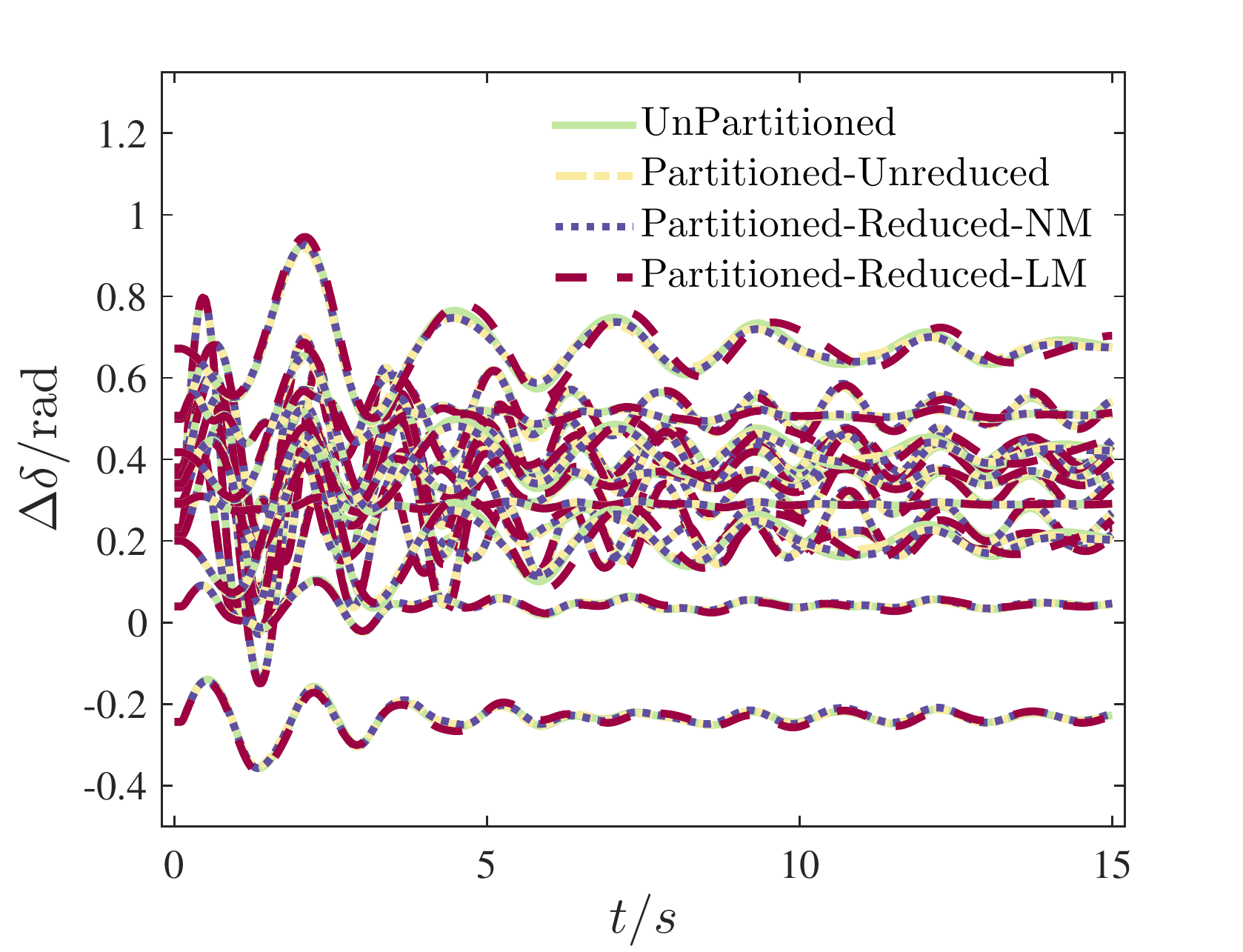}
\caption{Comparison of rotor angles of the study area for proposed method.}
\label{angle}
\end{figure}

\begin{figure}[!t]
\centering
\includegraphics[width=3.4in]{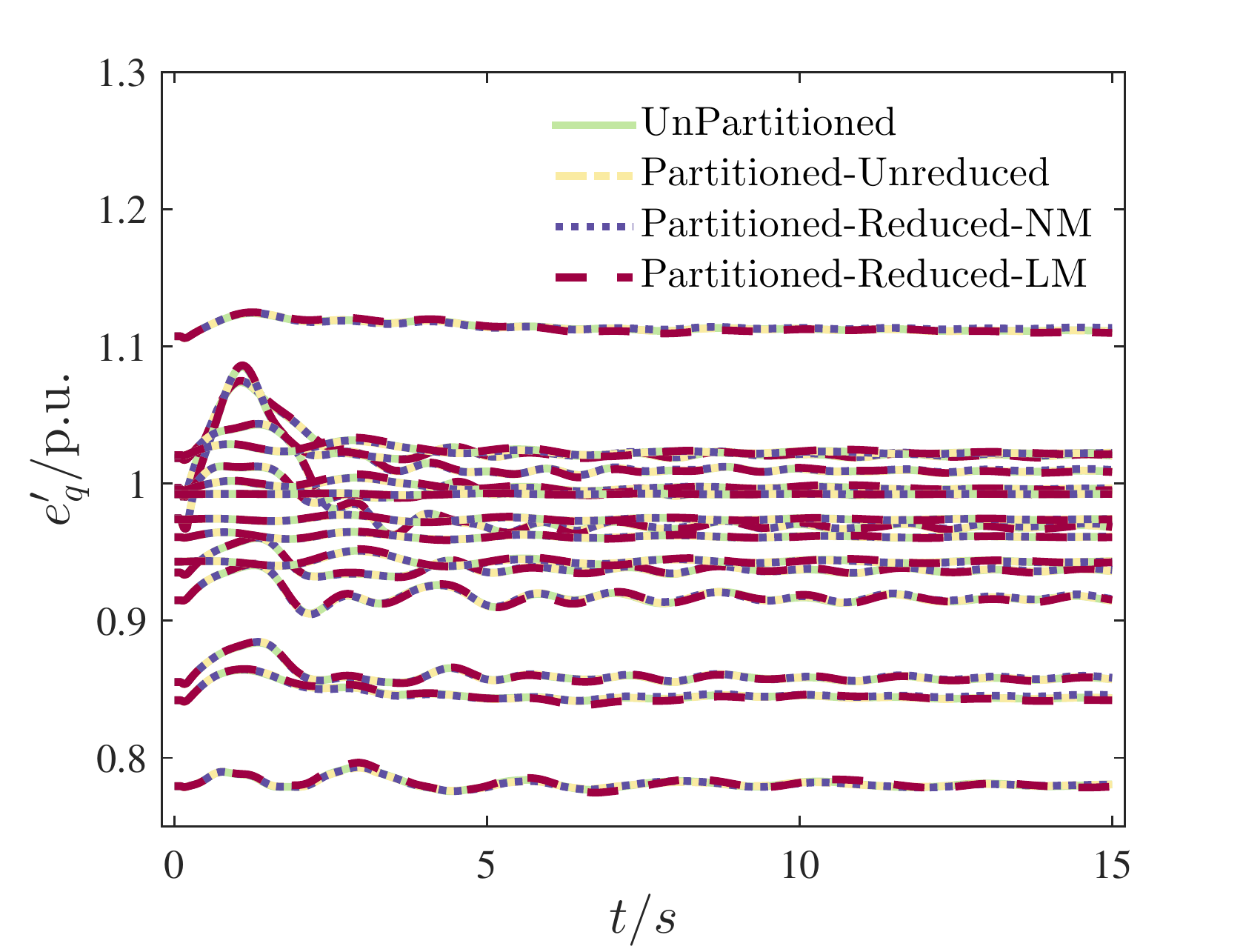}
\caption{Comparison of $e'_q$ of the study area for proposed method.}
\label{eqprime}
\end{figure}

There are $G_s=16$ generators in the study area whose states are of direct interest. 
In Figs. \ref{angle} and \ref{eqprime}, we present results for rotor angle and transient voltage along $q$-axis of the study area 
when the proposed model reduction and the model reduction in \cite{liu} are performed for the external area. 
For rotor angles, generator 13 in the study area is used as the reference. 
We can see that the results for ``Partitioned-Reduced-NM'' are closer to those for the ``UnPartitioned'' and ``Partitioned-Unreduced'' methods, compared with those for ``Partitioned-Reduced-LM''.

In order to quantify the accuracy of the model reduction methods, we define the following index:
\begin{align} \label{index}
\epsilon_s = \sqrt{\frac{\sum\limits_{i=1}^{N}\sum\limits_{t=1}^{T_s}(x_{i,t}^{\textrm{red}}-x_{i,t}^{\textrm{unred}})^2}{N\,T_s}}
\end{align}
where $x$ is one type of states and can be $\delta$, $\omega$, $e'_q$, $e'_d$, $V_R$, $E_{fd}$, or $R_f$;
$x_{i,t}^{\textrm{red}}$ is the simulated $i$th state for ``Partitioned-Reduced-NM'' or ``Partitioned-Reduced-LM'' method and $x_{i,t}^{\textrm{unred}}$ is the $i$th state from simulations without doing model reduction, both for time step $t$; $N$ is the number of trajectories to be compared, and here $N=G_s$, and $T_s$ is the total number of time steps. When we compare results from methods doing model reduction with ``UnPartitioned'' or ``Partitioned-Unreduced'' method, $\epsilon_s$ will be separately denoted by $\epsilon_s^1$ or $\epsilon_s^2$, which are listed in Table \ref{error_study}. It can be seen that for all types of state variables the defined indices for the proposed method are much smaller than those for the method in \cite{liu}.

\begin{table}[!t]
\renewcommand{\arraystretch}{1.4}
\caption{Simulation Accuracy for States in the Study Area}
\label{error_study}
\centering 
\begin{tabu}{ccccc}
\hline
     \multirow{3}{*}{Variable}  & \multicolumn{2}{c}{$\epsilon_s^1$} & \multicolumn{2}{c}{$\epsilon_s^2$} \\ 
     & \tabincell{c}{Partitioned- \\  Reduced-NM}  & \tabincell{c}{Partitioned- \\  Reduced-LM}  
    & \tabincell{c}{Partitioned- \\  Reduced-NM}  & \tabincell{c}{Partitioned- \\  Reduced-LM}  \\
    \midrule
    $\delta$ & $4.7 \times 10^{-2}$ & $2.3 \times 10^{-1}$    &   $6.0 \times 10^{-2}$   & $2.1 \times 10^{-1}$  \\
    $\omega$ & $2.3 \times 10^{-2}$ & $5.3 \times 10^{-2}$ & $1.9 \times 10^{-3}$ & $5.6 \times 10^{-2}$ \\
    $e'_q$ & $5.0 \times 10^{-4}$ & $8.4 \times 10^{-4}$ & $4.9 \times 10^{-4}$ & $8.7 \times 10^{-4}$ \\
    $e'_d$ & $3.7 \times 10^{-4}$ & $6.2 \times 10^{-4}$ & $3.1 \times 10^{-4}$ & $6.9 \times 10^{-4}$ \\
    $V_R$ & $1.7 \times 10^{-2}$ & $2.7 \times 10^{-2}$ & $6.0 \times 10^{-3}$ & $2.3 \times 10^{-2}$ \\
    $E_{fd}$ & $5.0 \times 10^{-3}$ & $1.0 \times 10^{-2}$ & $3.4 \times 10^{-3}$ & $1.1 \times 10^{-2}$ \\
    $R_f$ & $2.5 \times 10^{-3}$ & $4.9 \times 10^{-3}$ & $2.2 \times 10^{-3}$ & $5.1 \times 10^{-3}$ \\ 
\hline
\end{tabu}
\end{table}

\subsection{Results for Boundary Buses} \label{bou_result}

\begin{figure}[!t]
\centering
\includegraphics[width=3.4in]{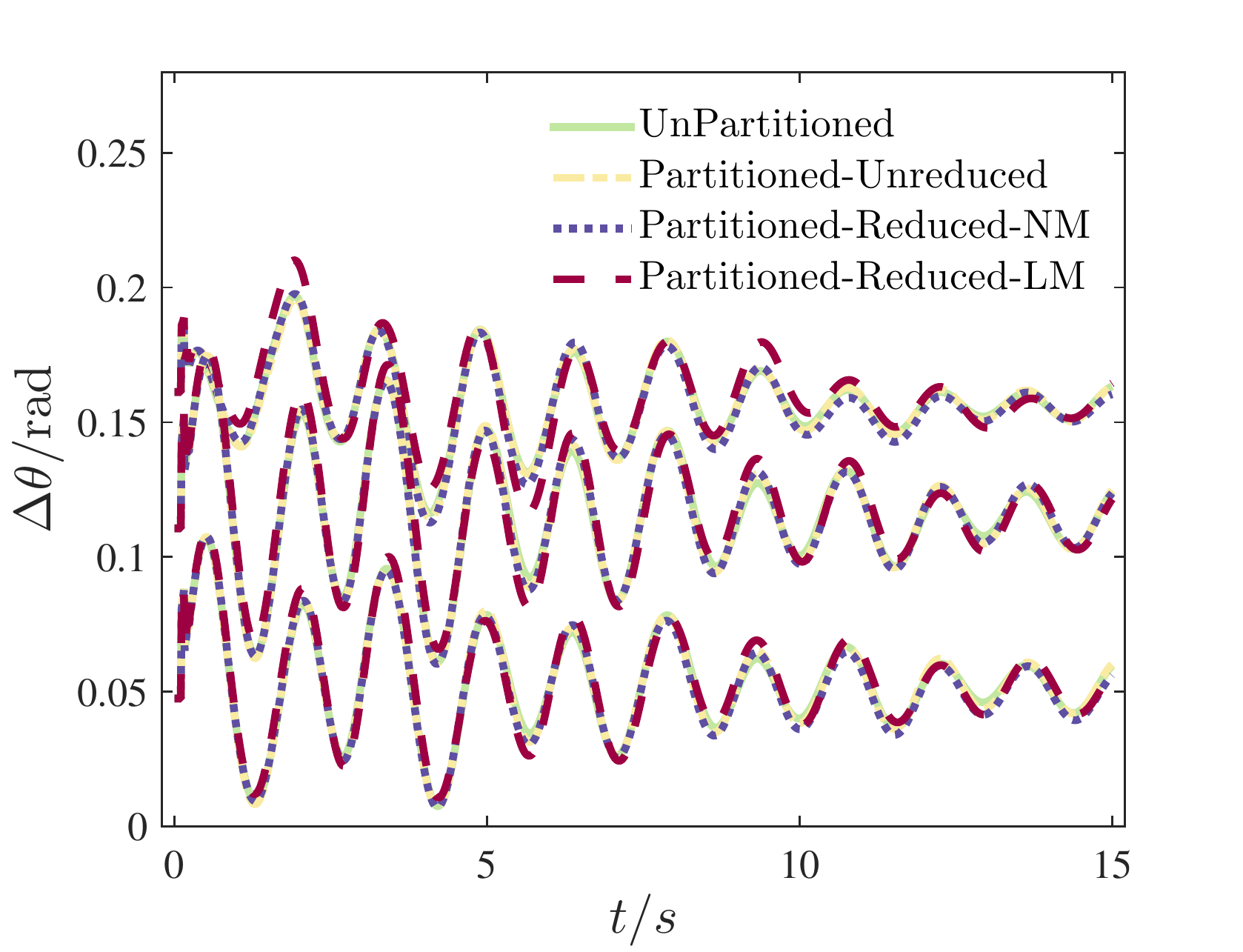}
\caption{Comparison of phase angle differences between boundary buses for proposed method and method in \cite{liu}.}
\label{angle_difference}
\end{figure}

The results for the phase angle differences between boundary buses for both model reduction methods are shown in Fig. \ref{angle_difference}.
It can be seen that the phase angle differences from the proposed method are very close to those from 
the ``UnPartitioned'' and ``Partitioned-Unreduced'' methods, while for the reduction method in \cite{liu} the differences are more obvious. 

A similar index to that in (\ref{index}) can be defined (denoted by $\epsilon_b^1$ and $\epsilon_b^2$, respectively, for comparison with the ``UnPartitioned'' and ``Partitioned-Unreduced'' methods) for the boundary buses for which $x$ is a type of variable for boundary buses and can be voltage magnitude ($V_s$ or $V_e$) or phase angles ($\theta_s$ or $\theta_e$), $N=3$ for our case is the number of boundary buses in each area. The defined indices for the proposed method can be much smaller than those for the method in \cite{liu}, as in Table \ref{error_boundary}.

\begin{table}[!t]
\renewcommand{\arraystretch}{1.4}
\caption{Simulation Accuracy for Boundary Buses}
\label{error_boundary}
\centering 
\begin{tabu}{ccccc}
\hline
    \multirow{3}{*}{Variable}  & \multicolumn{2}{c}{$\epsilon_b^1$} & \multicolumn{2}{c}{$\epsilon_b^2$}\\
     & \tabincell{c}{Partitioned- \\  Reduced-NM}  & \tabincell{c}{Partitioned- \\  Reduced-LM}  
    & \tabincell{c}{Partitioned- \\  Reduced-NM}  & \tabincell{c}{Partitioned- \\  Reduced-LM}  \\
    \midrule
    $V_s$ & $8.4 \times 10^{-4}$ & $1.5 \times 10^{-3}$    &   $7.6 \times 10^{-4}$   & $1.5 \times 10^{-3}$  \\
    $V_e$ & $2.4 \times 10^{-3}$ & $2.1 \times 10^{-3}$ & $2.4 \times 10^{-3}$ & $1.8 \times 10^{-3}$ \\
    $\theta_s$ & $4.7 \times 10^{-2}$ & $2.3 \times 10^{-1}$ & $6.1 \times 10^{-2}$ & $2.1 \times 10^{-1}$ \\
    $\theta_e$ & $4.8 \times 10^{-2}$ & $2.3 \times 10^{-1}$ & $6.2 \times 10^{-2}$ & $2.1 \times 10^{-1}$ \\
\hline
\end{tabu}
\end{table}

\subsection{Sensitivity Analysis for Empirical Covariance Calculation}

Here, we perform sensitivity analysis about how the empirical covariance calculation influences the accuracy of model reduction. 
Firstly, the $M_0$ in (\ref{Mc}) and (\ref{Mo}) chosen as a linearly scaled set in Section \ref{para_set} 
can also be chosen to be a geometrically scaled set as $\{0.125, 0.25, 0.5, 1.0\}$. 
Secondly, the $k_u$ and $k_x$ determined in Section \ref{para_set} can be scaled by a factor, such as 1/2 or 2. 

Therefore, we have six ways of setting $M^c$ and $M^o$, which are 
linearly scaled (LS), linearly scaled with halved $k_u$ and $k_x$ (LS-Half), linearly scaled with doubled $k_u$ and $k_x$ (LS-Double), 
geometrically scaled (GS), geometrically scaled with halved $k_u$ and $k_x$ (GS-Half), and geometrically scaled with doubled $k_u$ and $k_x$ (GS-Double). 
Then the model reduction can be performed for the external area separately based on the calculated empirical covariances for each $M^c$ and $M^o$. 
In Tables \ref{error_study_sensi_lin}--\ref{error_boundary_sensi_geo}, 
we list the simulation accuracy index $\epsilon_s^1$ and $\epsilon_b^1$ defined in Sections \ref{study_result} and \ref{bou_result} 
and for brevity we do not present results for $\epsilon_s^2$ or $\epsilon_b^2$.  
From these table, we can see that the simulation accuracy index $\epsilon_s^1$ and $\epsilon_b^1$ are very similar for different ways of setting 
$M^c$ and $M^o$, indicating that the model reduction is not sensitive to the choice of $M^c$ and $M^o$. 

\begin{table}[!t]
\renewcommand{\arraystretch}{1.4}
\caption{Simulation Accuracy for States in the Study Area for Empirical Covariances with Linear Scale ($\epsilon_s^1$)}
\label{error_study_sensi_lin}
\centering 
\begin{tabu}{cccc}
\hline
     Variable  & \tabincell{c}{LS}  & \tabincell{c}{LS-Half} & \tabincell{c}{LS-Double}  \\
    \midrule
    $\delta$ & $4.7 \times 10^{-2}$ & $4.1 \times 10^{-2}$    &   $2.8 \times 10^{-2}$   \\
    $\omega$ & $2.3 \times 10^{-2}$ & $2.2 \times 10^{-2}$ & $2.2 \times 10^{-2}$  \\
    $e'_q$ & $5.0 \times 10^{-4}$ & $4.5 \times 10^{-4}$ & $5.6 \times 10^{-4}$  \\
    $e'_d$ & $3.7 \times 10^{-4}$ & $3.4 \times 10^{-4}$ & $4.2 \times 10^{-4}$  \\
    $V_R$ & $1.7 \times 10^{-2}$ & $1.6 \times 10^{-2}$ & $1.8 \times 10^{-2}$  \\
    $E_{fd}$ & $5.0 \times 10^{-3}$ & $4.8 \times 10^{-3}$ & $6.1 \times 10^{-3}$  \\
    $R_f$ & $2.5 \times 10^{-3}$ & $2.3 \times 10^{-3}$ & $3.1 \times 10^{-3}$  \\ 
\hline
\end{tabu}
\end{table}

\begin{table}[!t]
\renewcommand{\arraystretch}{1.4}
\caption{Simulation Accuracy for Boundary Buses for Empirical Covariances with Linear Scale ($\epsilon_b^1$)}
\label{error_boundary_sensi_lin}
\centering 
\begin{tabu}{cccc}
\hline
     Variable  & \tabincell{c}{LS}  & \tabincell{c}{LS-Half}  & \tabincell{c}{LS-Double}  \\
    \midrule
    $V_s$  & $8.4 \times 10^{-4}$ & $7.7 \times 10^{-4}$    &   $1.0 \times 10^{-3}$   \\
    $V_e$  & $2.4 \times 10^{-3}$ & $2.2 \times 10^{-3}$ & $2.8 \times 10^{-3}$  \\
    $\theta_s$ & $4.7 \times 10^{-2}$ & $4.2 \times 10^{-2}$ & $2.7 \times 10^{-2}$  \\
    $\theta_e$ & $4.8 \times 10^{-2}$ & $4.2 \times 10^{-2}$ & $2.8 \times 10^{-2}$  \\
\hline
\end{tabu}
\end{table}

\begin{table}[!t]
\renewcommand{\arraystretch}{1.4}
\caption{Simulation Accuracy for States in the Study Area for Empirical Covariances with Geometric Scale ($\epsilon_s^1$)}
\label{error_study_sensi_geo}
\centering 
\begin{tabu}{cccc}
\hline
     Variable  & \tabincell{c}{GS}  & \tabincell{c}{GS-Half}  & \tabincell{c}{GS-Double}  \\
    \midrule
    $\delta$ & $4.4 \times 10^{-2}$ & $4.0 \times 10^{-2}$    &   $3.0 \times 10^{-2}$   \\
    $\omega$ & $2.3 \times 10^{-2}$ & $2.2 \times 10^{-2}$ & $2.2 \times 10^{-2}$  \\
    $e'_q$ & $4.7 \times 10^{-4}$ & $4.4 \times 10^{-4}$ & $4.4 \times 10^{-4}$  \\
    $e'_d$ & $3.5 \times 10^{-4}$ & $3.4 \times 10^{-4}$ & $3.5 \times 10^{-4}$  \\
    $V_R$ & $1.7 \times 10^{-2}$ & $1.6 \times 10^{-2}$ & $1.8 \times 10^{-2}$  \\
    $E_{fd}$ & $4.9 \times 10^{-3}$ & $4.8 \times 10^{-3}$ & $5.5 \times 10^{-3}$  \\
    $R_f$ & $2.4 \times 10^{-3}$ & $2.3 \times 10^{-3}$ & $2.5 \times 10^{-3}$ \\ 
\hline
\end{tabu}
\end{table}

\begin{table}[!t]
\renewcommand{\arraystretch}{1.4}
\caption{Simulation Accuracy for Boundary Buses for Empirical Covariances with Geometric Scale ($\epsilon_b^1$)}
\label{error_boundary_sensi_geo}
\centering 
\begin{tabu}{cccc}
\hline
     Variable  & \tabincell{c}{GS}  & \tabincell{c}{GS-Half}  & \tabincell{c}{GS-Double}  \\
    \midrule
    $V_s$ & $8.1 \times 10^{-4}$ & $7.6 \times 10^{-4}$    &   $8.5 \times 10^{-4}$   \\
    $V_e$ & $2.3 \times 10^{-3}$ & $2.1 \times 10^{-3}$ & $2.2 \times 10^{-3}$  \\
    $\theta_s$ & $4.5 \times 10^{-2}$ & $4.1 \times 10^{-2}$ & $3.0 \times 10^{-2}$  \\
    $\theta_e$ & $4.5 \times 10^{-2}$ & $4.1 \times 10^{-2}$ & $3.0 \times 10^{-2}$  \\
\hline
\end{tabu}
\end{table}

\subsection{Efficiency}

The calculation times, $t_{\textrm{total}}$, for simulating 15 seconds by different methods are listed in Table \ref{eff}. 
Since the times for different ways of setting $M^c$ and $M^o$ are similar, we only list the time for linearly scaled $M_0$.
It is seen that our proposed model reduction method can improve the calculation efficiency of dynamic simulation 
and help achieve faster than real-time simulation. 
Also, the efficiency of our model reduction method based on a nonlinear model is similar to that for the balanced truncation method in \cite{liu} based on a linearized model.

\begin{table}[!t]
\renewcommand{\arraystretch}{1.4}
\caption{Total Time in Second for Simulating 15 Seconds}
\label{eff}
\centering
\begin{tabu}{cccc}
\hline
 UnPartitioned & \tabincell{c}{Partitioned- \\  UnReduced} & \tabincell{c}{Partitioned- \\  Reduced-NM} & \tabincell{c}{Partitioned- \\  Reduced-LM} \\
\hline
26.99 &  23.16 & 14.44  & 13.90 \\
\hline
\end{tabu}
\end{table}

To clearly identify the bottleneck of the proposed method and that in \cite{liu}, in Table \ref{eff_split} we list 
the calculation time for the three steps in Section \ref{simu_whole}. 
Here, $t_s$, $t_e$, and $t_b$ are the time for simulating the study area, the external area, and updating the boundary buses, respectively.
For both model reduction methods, most calculation time is for simulating the detailed modeled study area. 
The calculation time of simulating the external area for nonlinear model reduction is a little higher than that based on a
linearized model, which explains why the $t_{\textrm{total}}$ for the nonlinear model reduction is a little higher. 

Note that the first two steps in Section \ref{simu_whole} are decoupled and can be calculated in parallel, which can further improve the simulation efficiency. 
Then the total calculation time will be $t'_{\textrm{total}}=\max \{t_e,t_s\} + t_b$, which is also listed in Table \ref{eff_split}. 
The simulation speedup finally achieves $23.16/12.30 \cong 1.88$ and the simulation is $15/12.30 \cong 1.22$ times faster than real time.

In this test case, if the first two steps in Section \ref{simu_whole} are calculated in parallel, 
the advantage of the model reduction methods over the ``Partitioned-Unreduced'' method is not obvious. 
This is because the external area in our test case is not significantly larger than the study area. 
In the case that the external area is much larger than the study area, we will have
\begin{align}
\frac{t'_{\textrm{total}}(\textrm{Par})}{t'_{\textrm{total}}(\textrm{Red})}&=\frac{\max\{t_s(\textrm{Par}),t_e(\textrm{Par})\}+t_b(\textrm{Par})}{\max \{t_s(\textrm{Red})+t_e(\textrm{Red})\}+t_b(\textrm{Red})} \notag \\
&=\frac{t_e(\textrm{Par})+t_b(\textrm{Par})}{t_e(\textrm{Red})+t_b(\textrm{Red})} \cong \frac{t_e(\textrm{Par})}{t_e(\textrm{Red}))}
\end{align}
where ``Par'' represents the ``Partitioned-Unreduced'' method and ``Red'' indicates the model reduction methods, either nonlinear or linear model reduction. 
The speedup for the model reduction methods compared with the ``Partitioned-Unreduced'' method can achieve $t_e(\textrm{Par})/t_e(\textrm{Red})$. 
If we assume the speedup for the external area simulation for larger external areas is the same as that in our test case, 
then the speedup can be $10.57/2.14\cong4.94$ or $10.57/1.58 \cong 6.69$ for the proposed nonlinear model reduction and the method in \cite{liu} based on a linearized model, respectively.

\begin{table}[!t]
\renewcommand{\arraystretch}{1.4}
\caption{Time for the Three Steps in Section \ref{simu_whole}}
\label{eff_split}
\centering
\begin{tabu}{c|ccc}
\hline
 Method & \tabincell{c}{Partitioned- \\  UnReduced} & \tabincell{c}{Partitioned- \\  Reduced-NM} & \tabincell{c}{Partitioned- \\  Reduced-LM} \\
\hline
$t_s$\,(s) & 10.54 & 10.24 &  10.28  \\
$t_e$\,(s) & 10.57 & 2.14 &  1.58  \\
$t_b$\,(s) & 2.05 & 2.06 &  2.04  \\
$t'_{\textrm{total}}$\,(s) & 12.62 & 12.30 &  12.32  \\
\hline
\end{tabu}
\end{table}

\section{Conclusion} \label{conclusion}

In this paper, a nonlinear power system model reduction method is proposed by balancing of the empirical controllability and observability covariances. 
Compared with the balanced truncation method based on a linearized model, the proposed model reduction method can guarantee higher accuracy for simulated state trajectory, mainly because the empirical covariances are defined using the original system model and can thus reflect the controllability and observability of the full nonlinear dynamics in the given domain. 

The proposed method is validated on a test system comprised of a 16-machine 68-bus system as the study area and an IEEE 50-machine 145-bus system as the external area. 
The results show that by using the proposed model reduction method the simulation efficiency is greatly improved and 
at the same time the obtained state trajectories are close to those for directly simulating the whole system and for partitioning the system while not performing reduction. 
By contrast, for the balanced truncation method based on a linearized model when using the balancing transformation method in Section \ref{reduction}, 
the simulation accuracy is lower but is still acceptable, and the calculation efficiency is similar to that of our proposed model reduction method. 
However, when the balancing transformation method from \cite{laub} is applied for the balanced truncation method based on a linearized model, as in \cite{liu}, 
the simulation cannot proceed, which is mainly because that balancing transformation is not applicable to systems that are not completely controllable and observable.

By solving the differential equations in the study area and the external area in parallel, in our test case the speedup compared with the ``UnPartitioned'' method finally achieves 1.88 and the simulation is 1.22 times faster than real time. When the external system is much larger than the study area, the speedup of the proposed method compared with the ``Partitioned-Unreduced'' method can achieve 4.94.
It is also shown that the proposed model reduction method is not sensitive to the choice of the matrices for calculating the empirical controllability and observability covariances.

\appendices

\section{Model for Study Area} \label{appA}

For the study area, the fast sub-transient dynamics and saturation effects are ignored and the generator is described by the two-axis transient model with IEEE Type DC1 excitation system \cite{sauer}:
\begin{subnumcases}{\label{gen model}}
\dot{\delta_i}=\omega_i-\omega_0 \\
\dot{\omega}_i=\frac{\omega_0}{2H_i}\Big(T_{\mathrm{m}i}-T_{\mathrm{e}i}-\frac{K_{\mathrm{D}i}}{\omega_0}(\omega_i-\omega_0)\Big) \\
\dot{e}'_{\mathrm{q}i}=\frac{1}{T'_{\mathrm{d0}i}}\Big(E_{\mathrm{fd}i}-e'_{\mathrm{q}i}-(x_{\mathrm{d}i}-x'_{\mathrm{d}i})\,i_{\mathrm{d}i}\Big) \\
\dot{e}'_{\mathrm{d}i}=\frac{1}{T'_{\mathrm{q0}i}}\Big(-e'_{\mathrm{d}i}+(x_{\mathrm{q}i}-x'_{\mathrm{q}i})\,i_{\mathrm{q}i}\Big) \\
\dot{V}_{\mathrm{R}i}=\frac{1}{T_{\mathrm{A}i}}(-V_{\mathrm{R}i}+K_{\mathrm{A}i} V_{\mathrm{A}i}) \\
\dot{E}_{\mathrm{fd}i}=\frac{1}{T_{\mathrm{E}i}}(V_{\mathrm{R}i}-K_{\mathrm{E}i}E_{\mathrm{fd}i}-S_{\mathrm{E}i}) \\
\dot{R}_{\mathrm{f}i}=\frac{1}{T_{\mathrm{F}i}}(-R_{\mathrm{f}i}+E_{\mathrm{fd}i})
\end{subnumcases}
where $i$ is the generator serial number, $\delta_i$ is rotor angle, $\omega_i$ is rotor speed in rad/s, and $e'_{\mathrm{q}i}$ and $e'_{\mathrm{d}i}$ are transient voltage along $\mathrm{q}$ and $\mathrm{d}$ axes; $i_{\mathrm{q}i}$ and $i_{\mathrm{d}i}$ are stator currents at $\mathrm{q}$ and $\mathrm{d}$ axes; $V_{\mathrm{R}i}$ is regulator output voltage, $E_{\mathrm{fd}i}$ is excitation output voltage, $R_{\mathrm{f}i}$ is stabilizing transformer state variable; $T_{\mathrm{m}i}$ is mechanical torque, $T_{\mathrm{e}i}$ is electric air-gap torque; $\omega_0$ is the rated value of angular frequency, $H_i$ is inertia constant, and $K_{\mathrm{D}i}$ is damping factor; $T'_{\mathrm{q0}i}$ and $T'_{\mathrm{d0}i}$ are  open-circuit time constants, $x_{\mathrm{q}i}$ and $x_{\mathrm{d}i}$ are synchronous reactance, and $x'_{\mathrm{q}i}$ and $x'_{\mathrm{d}i}$ are transient reactance, respectively, at the $\mathrm{q}$ and $\mathrm{d}$ axes; $T_{\mathrm{A}i}$ is voltage regulator time constant, $T_{\mathrm{E}i}$ is exciter time constant, $T_{\mathrm{F}i}$ is stabilizer time constant, $K_{\mathrm{A}i}$ is voltage regulator gain, and $K_{\mathrm{E}i}$ is exciter constant.

The load buses in $\mathcal{B}_{s,\textrm{ZIP}}$ are modeled as a combination of constant impedance, constant current, and constant power (also called non-conforming load, as in \cite{pst}) as
\begin{align}
P_i&= P_{0,i}\Bigg( p_1\bigg(\frac{|\tilde{V}_{\mathrm{nc},i}|}{V_{\mathrm{nc0},i}}\bigg)^2 + p_2 \bigg(\frac{|\tilde{V}_{\mathrm{nc},i}|}{V_{\mathrm{nc0},i}}\bigg) + p_3 \Bigg)  \\
Q_i&= Q_{0,i}\Bigg( q_1\bigg(\frac{|\tilde{V}_{\mathrm{nc},i}|}{|\tilde{V}_{\mathrm{nc0},i}|}\bigg)^2 + q_2 \bigg(\frac{|\tilde{V}_{\mathrm{nc},i}|}{|\tilde{V}_{\mathrm{nc0},i}|}\bigg) + q_3 \Bigg) 
\end{align}
where $P_{i}$ and $Q_{i}$ are the active and reactive power at load bus $i$, $P_{0,i}$ and $Q_{0,i}$ are the initial active and reactive power at load bus $i$, $p_1$, $p_2$, and $p_3$ are proportions of constant active impedance load, constant active current load, and constant active power load, $q_1$, $q_2$, and $q_3$ are proportions of constant reactive impedance load, constant reactive current load, and constant reactive power load, and there is $p_1+p_2+p_3=1$ and $q_1+q_2+q_3=1$, $\tilde{V}_{\mathrm{nc},i}$ and $\tilde{V}_{\mathrm{nc0},i}$ are the complex voltage and initial complex voltage at load bus $i$. The other load buses that do not belong to $\mathcal{B}_{s,\textrm{ZIP}}$ are modeled as constant impedance.

The input and output are, respectively, the voltage magnitude and phase angles of the boundary buses in external area and study area. 
The boundary buses in the external area are treated as generators with a classical second-order model and very large inertia constant, 
which can be described by the first two equations in (\ref{gen model}). 
The voltage magnitude and phase angles of the boundary buses in external area are respectively used as the $e'_\mathrm{q}$ and $\delta$ of the equivalent generator, for which 
$\omega=\omega_0$ and $e'_\mathrm{d}=0$. 
The dynamic model (\ref{gen model}) can be rewritten in a general state space form in (\ref{n1}) and
the state vector $\boldsymbol{x}_s$, input vector $\boldsymbol{u}_s$, and output vector $\boldsymbol{y}_s$
can be written as
\begin{subequations}
\begin{align}
\boldsymbol{x}_s &= \big[\boldsymbol{\delta}_s^\top \;\; \boldsymbol{\omega}_s^\top \;\; \boldsymbol{e'_{\mathrm{q}}}_s^\top \;\; \boldsymbol{e'_{\mathrm{d}}}_s^\top \;\; \boldsymbol{V_\mathrm{R}}_s^\top \;\; \boldsymbol{E_{\mathrm{fd}}}_s^\top \;\; \boldsymbol{R_\mathrm{f}}_s^\top\big]^\top \\
\boldsymbol{u}_s &= \big[\boldsymbol{V}_e^\top \;\; \boldsymbol{\theta}_e^\top\big]^\top \\
\boldsymbol{y}_s &= \big[\boldsymbol{V}_s^\top \;\; \boldsymbol{\theta}_s^\top\big]^\top.
\end{align}
\end{subequations}

The $i_{\mathrm{q}i}$, $i_{\mathrm{d}i}$, $T_{\mathrm{e}i}$, $V_{\mathrm{A}i}$, and $S_{\mathrm{E}i}$ in (\ref{gen model}) can be written as functions of $\boldsymbol{x}_s$ and $\boldsymbol{u}_s$ (note that for boundary bus $b_i^e$ in external area, the generator number is $g_i^e$ and there are $e'_{\mathrm{q}g_i^e}=V_{eb_i^e}$, $e'_{\mathrm{d}g_i^e}=0$, and $\delta_{g_i^e}=\theta_{b_i^e}$):
\begin{subequations} \label{temp}
\begin{align}
&\it \Psi_{\mathrm{R}i}=e'_{\mathrm{d}i}\sin\delta_i+e'_{\mathrm{q}i}\cos\delta_i \label{te1} \\
&\it \Psi_{\mathrm{I}i}=e'_{\mathrm{q}i}\sin\delta_i-e'_{\mathrm{d}i}\cos\delta_i \\
& I_{\mathrm{t}i}=\overline{\boldsymbol{Y}}_{\mathrm{g},i}(\boldsymbol{\it \Psi}_{\mathbf{R}}+j \boldsymbol{\it \Psi}_{\mathbf{I}}) + \overline{\boldsymbol{Y}}_{\mathrm{gnc},i} \tilde{\boldsymbol{V}}_{\mathrm{nc}} \label{35c} \\ 
& i_{\mathrm{R}i}= \operatorname{Re}(I_{\mathrm{t}i}) \\
& i_{\mathrm{I}i}= \operatorname{Im}(I_{\mathrm{t}i}) \\
& i_{\mathrm{q}i} = \frac{S_\mathrm{B}}{S_{\mathrm{N}i}}(i_{\mathrm{I}i}\sin\delta_i+i_{\mathrm{R}i}\cos\delta_i) \\
& i_{\mathrm{d}i} = \frac{S_\mathrm{B}}{S_{\mathrm{N}i}}(i_{\mathrm{R}i}\sin\delta_i-i_{\mathrm{I}i}\cos\delta_i) \\
& e_{\mathrm{q}i}=e'_{\mathrm{q}i}-x'_{\mathrm{d}i}i_{\mathrm{d}i} \\
& e_{\mathrm{d}i}=e'_{\mathrm{d}i}+x'_{\mathrm{q}i}i_{\mathrm{q}i} \\
& P_{\mathrm{e}i} = e_{\mathrm{q}i}i_{\mathrm{q}i}+e_{\mathrm{d}i}i_{\mathrm{d}i} \\
& T_{\mathrm{e}i} = \frac{S_\mathrm{B}}{S_{\mathrm{N}i}} P_{\mathrm{e}i} \label{te2} \\
& V_{\mathrm{FB}i}=\frac{K_{\mathrm{F}i}}{T_{\mathrm{F}i}}(E_{\mathrm{fd}i}-R_{\mathrm{f}i}) \\
& V_{\mathrm{TR}i}=\sqrt{{e_{\mathrm{d}i}}^2+{e_{\mathrm{q}i}}^2} \\
& V_{\mathrm{A}i}= -V_{\mathrm{FB}i}+ \mathrm{exc}_i^3 - V_{\mathrm{TR}i} \\
& S_{\mathrm{E}i}= \mathrm{exc}_i^1\, e^{\mathrm{exc}_i^2 |E_{\mathrm{fd}i}|} \textrm{sgn}(E_{\mathrm{fd}i})
\end{align}
\end{subequations}
where $\it \Psi_i=\Psi_{\mathrm{R}i}+j\Psi_{\mathrm{I}i}$ is the voltage source, $\boldsymbol{\it \Psi}=\boldsymbol{\it \Psi_\mathrm{R}}+j \boldsymbol{\it \Psi_\mathrm{I}}$ is the column vector of all generators' voltage sources, $e_{\mathrm{q}i}$ and $e_{\mathrm{d}i}$ are the terminal voltage at $\mathrm{q}$ and $\mathrm{d}$ axes, 
$\overline{\boldsymbol{Y}}_{\mathrm{g},i}$ is the $i$th row of the reduced admittance matrix connecting the generator current injections to the
internal generator voltages (including boundary buses in external area) $\boldsymbol{\overline{Y}}_\mathrm{g}$, 
and $\overline{\boldsymbol{Y}}_{\mathrm{gnc},i}$ is the $i$th row of the reduced admittance matrix which gives the generator currents
due to the voltages at non-conforming loads $\boldsymbol{\overline{Y}}_{\mathrm{gnc}}$;
$P_{\mathrm{e}i}$ is the electrical active output power, and $S_\mathrm{B}$ and $S_{\mathrm{N}i}$ are the system base MVA and the base MVA for generator $i$;
$K_{\mathrm{F}i}$ is the stabilizer gain; $\mathrm{exc}_i^1$, $\mathrm{exc}_i^2$, and $\mathrm{exc}_i^3$ are internally set exciter constants; 
and $\textrm{sgn}(\cdot)$ is the signum function. 
The $\tilde{\boldsymbol{V}}_{\mathrm{nc}}$ in (\ref{35c}) is the complex voltages of the non-conforming load buses and can be obtained by solving the following nonlinear equations by Newton's method:
\begin{align} \label{vnc}
\overline{\boldsymbol{Y}}_{\mathrm{ncg}} \boldsymbol{\it \Psi} + \overline{\boldsymbol{Y}}_{\mathrm{nc}} \tilde{\boldsymbol{V}}_{\mathrm{nc}} = \tilde{\boldsymbol{I}}_{\mathrm{cc}} + \tilde{\boldsymbol{I}}_{\mathrm{cp}}
\end{align}
where $\overline{\boldsymbol{Y}}_{\mathrm{ncg}}$ is the reduced admittance matrix connecting non-conforming load current to machine internal voltages, $\overline{\boldsymbol{Y}}_{\mathrm{nc}}$ is the reduced admittance matrix of non-conforming loads, and $\tilde{\boldsymbol{I}}_{\mathrm{cc}}$ and $\tilde{\boldsymbol{I}}_{\mathrm{cp}}$ are current injections of the constant current and constant power components. $\tilde{\boldsymbol{I}}_{\mathrm{cc}}+\tilde{\boldsymbol{I}}_{\mathrm{cp}}$ is actually a function of $\tilde{\boldsymbol{V}}_{\mathrm{nc}}$. 
For $|\tilde{V}_{\mathrm{nc},i}| > 0.5$, it can be written as 
\begin{equation}
-\Bigg(\frac{p_3 P_{0,i}+p_2 P_{0,i}\frac{|\tilde{V}_{\mathrm{nc},i}|}{|\tilde{V}_{\mathrm{nc0},i}|} + j\Big(q_3 Q_{0,i} + q_2 Q_{0,i} \frac{|\tilde{V}_{\mathrm{nc},i}|}{|\tilde{V}_{\mathrm{nc0},i}|}\Big)}{\tilde{V}_{\mathrm{nc},i}}\Bigg)^* \notag
\end{equation}
while for $|\tilde{V}_{\mathrm{nc},i}| \le 0.5$ it is
\begin{equation}
-\bigg(\frac{p_3 P_{0,i}+j q_3 Q_{0,i}+p_2 P_{0,i}+j q_2 Q_{0,i}}{\tilde{V}_{\mathrm{nc0},i} \, \tilde{V}_{\mathrm{nc0},i}^*}\bigg)^* \, \tilde{V}_{\mathrm{nc},i} \notag
\end{equation}
where $(\cdot)^*$ is the complex conjugation.

The outputs can also be written as function of $\boldsymbol{x}_s$ and $\boldsymbol{u}_s$:
\begin{subequations}
\begin{align}
&\boldsymbol{\it \Psi}_s^{\mathrm{re}} = \boldsymbol{e'_\mathrm{d}}_{s} \sin \boldsymbol{\delta}_{s}  + \boldsymbol{e'_\mathrm{q}}_{s} \cos \boldsymbol{\delta}_{s}  \label{input1} \\
&\boldsymbol{\it \Psi}_s^{\mathrm{im}} =  \boldsymbol{e'_\mathrm{q}}_{s} \sin \boldsymbol{\delta}_{s}  - \boldsymbol{e'_\mathrm{d}}_{s} \cos \boldsymbol{\delta}_{s}  \\
&\boldsymbol{\it \Psi}_s^{\mathrm{state}} = \boldsymbol{\it \Psi}_s^{\mathrm{re}} + j \boldsymbol{\it \Psi}_s^{\mathrm{im}} \\
&\boldsymbol{\it \Psi}_s^{\mathrm{input}} = \boldsymbol{V}_{e}\, e^{\,j \boldsymbol{\theta}_{e}} \\
&\tilde{\boldsymbol{V}}_{s,\mathcal{B}_{s,\mathrm{ZIP}}}=\tilde{\boldsymbol{V}}_{\mathrm{nc}} \\
&\tilde{\boldsymbol{V}}_{s,\mathcal{B}_{s,\mathrm{ZIP}}^c}=\boldsymbol{R}_{\mathrm{gs}} [{\boldsymbol{\it \Psi}_s^{\mathrm{state}}}^\top \; {\boldsymbol{\it \Psi}_s^{\mathrm{input}}}^\top]^\top + \boldsymbol{R}_{\mathrm{nc}} \tilde{\boldsymbol{V}}_{\mathrm{nc}} \\
&\boldsymbol{V}_s=|\tilde{\boldsymbol{V}}_{s,\mathcal{B}_{s,\mathrm{bound}}}| \\
&\boldsymbol{\theta}_s=\arg(\tilde{\boldsymbol{V}}_{s,\mathcal{B}_{s,\mathrm{bound}}}). \label{input2}
\end{align}
\end{subequations}

\section{Model for External Area} \label{appB}

Both fourth-order and second-order generator model are used for the external area. 
In (\ref{gen model}), the generators with fourth-order model are described by the first four equations and ${V}_{\mathrm{R}i}$, ${E}_{\mathrm{fd}i}$, and ${R}_{\mathrm{f}i}$ are kept unchanged. 
The generators with second-order model are described only by the first two equations and $e'_{\mathrm{q}i}$, $e'_{\mathrm{d}i}$, ${V}_{\mathrm{R}i}$, ${E}_{\mathrm{fd}i}$, and ${R}_{\mathrm{f}i}$ are all kept unchanged. 
The input and output are respectively the voltage magnitude and phase angles of the boundary buses in study and external area. 
$T_{\mathrm{e}i}$ can be obtained by (\ref{te1})--(\ref{te2}) and the outputs can be calculated in a similar way to (\ref{input1})--(\ref{input2}) in Appendix \ref{appA}.
The dynamic model can be rewritten in the form (\ref{n1}) and the state vector, input vector, and output vector can be written as
\begin{subequations}
\begin{align}
\boldsymbol{x}_e & =\big[\boldsymbol{\delta}_e^\top \;\; \boldsymbol{\omega}_e^\top \;\; \boldsymbol{e'_{\mathrm{q}}}_e^\top \;\; \boldsymbol{e'_{\mathrm{d}}}_e^\top\big]^\top  \\
\boldsymbol{u}_e &= \big[\boldsymbol{V}_s^\top \;\; \boldsymbol{\theta}_s^\top\big]^\top \\
\boldsymbol{y}_e &= \big[\boldsymbol{V}_e^\top \;\; \boldsymbol{\theta}_e^\top\big]^\top.
\end{align}
\end{subequations}


\begin{IEEEbiography} [{\includegraphics[width=1in,height=1.25in,clip,keepaspectratio]{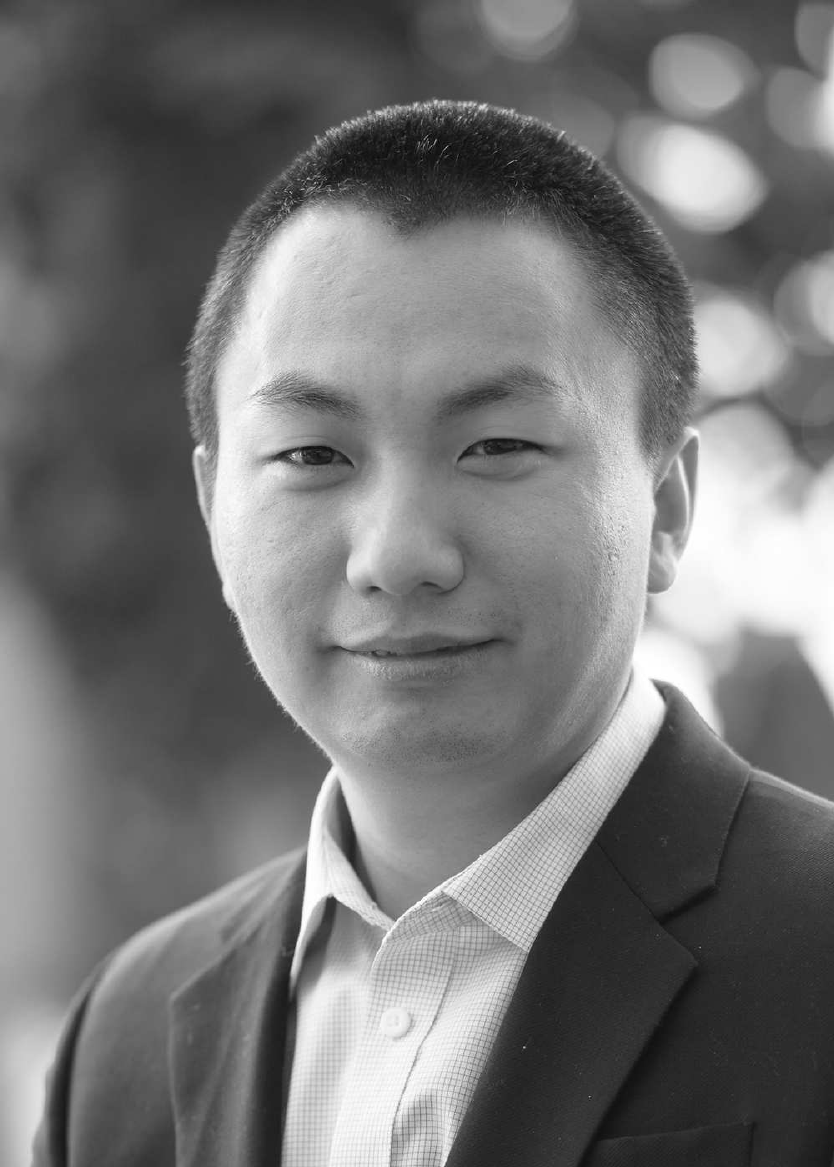}\vfill}]
{Junjian Qi} (S'12--M'13)
received the B.E. and Ph.D. degree both in electrical engineering from Shandong University, Shandong, China in 2008 and Tsinghua University, Beijing, China in 2013.

In Feb.--Aug. 2012 he was a Visiting Scholar at Iowa State University, Ames, IA, USA. During Sept. 2013--Jan. 2015 he was 
a Research Associate at Department of Electrical Engineering and Computer Science, University of Tennessee, Knoxville, TN, USA. 
Currently he is a Postdoctoral Appointee at the Energy Systems Division, Argonne National Laboratory, Argonne, IL, USA. 
His research interests include cascading blackouts, power system dynamics, state estimation, synchrophasors, and cybersecurity.
\end{IEEEbiography}

\begin{IEEEbiography} [{\includegraphics[width=1in,height=1.25in,clip,keepaspectratio]{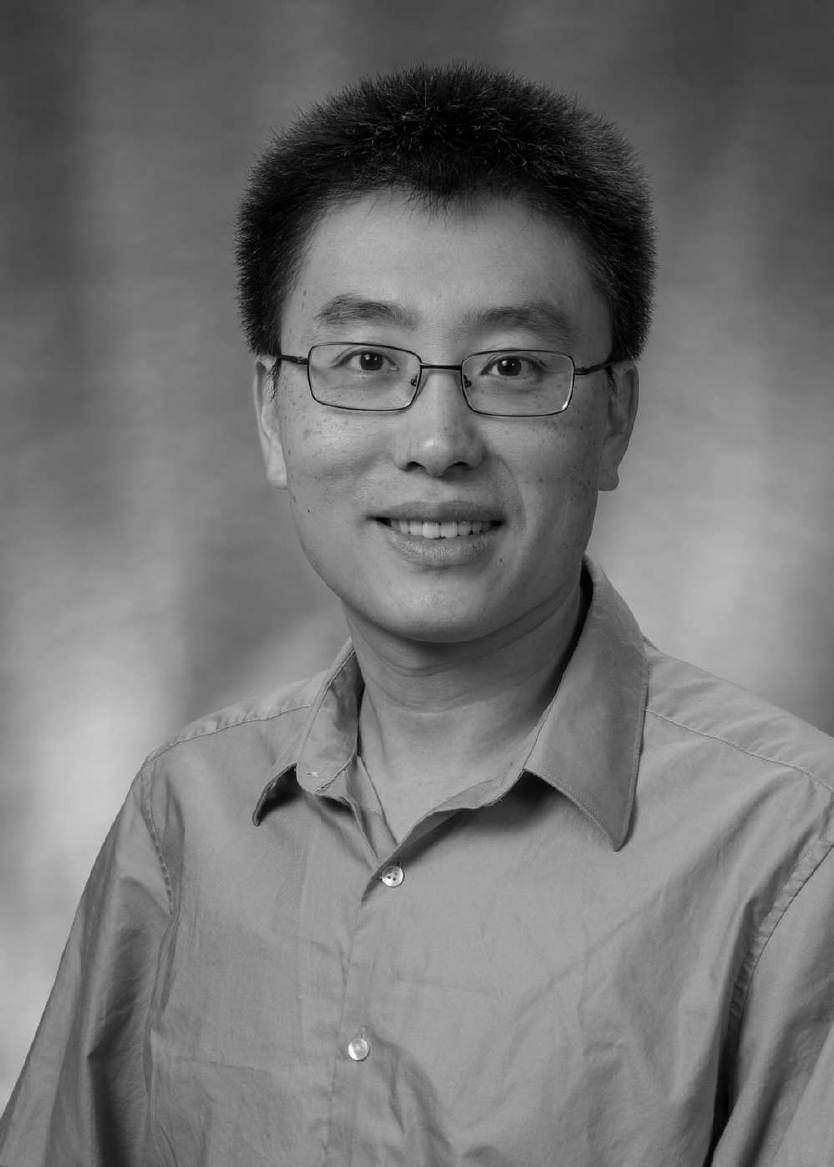}\vfill}]
{Jianhui Wang} (S'07--SM'12) 
received the Ph.D. degree in electrical engineering from Illinois Institute of Technology, Chicago, IL, USA, in 2007. 

Presently, he is the Section Lead for Advanced Power Grid Modeling at the Energy Systems Division at Argonne National Laboratory, Argonne, IL, USA.
Dr. Wang is the secretary of the IEEE Power \& Energy Society (PES) Power System Operations Committee. 

He is an Associate Editor of Journal of Energy Engineering and an editorial board member of Applied Energy. He is also an affiliate professor at Auburn University and an adjunct professor at University of Notre Dame. He has held visiting positions in Europe, Australia, and Hong Kong including a VELUX Visiting Professorship at the Technical University of Denmark (DTU). Dr. Wang is the Editor-in-Chief of the IEEE Transactions on Smart Grid and an IEEE PES Distinguished Lecturer. He is also the recipient of the IEEE PES Power System Operation Committee Prize Paper Award in 2015.
\end{IEEEbiography}


\begin{IEEEbiography} [{\includegraphics[width=1in,height=1.25in,clip,keepaspectratio]{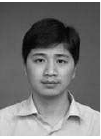}\vfill}]
{Hui Liu} (M'12)
received the M.S. degree in 2004 and the Ph.D. degree in 2007 from the School of Electrical Engineering at Guangxi University, China, both in electrical engineering.  

He was a Postdoctoral Fellow at Tsinghua University from 2011 to 2013 and was a staff at Jiangsu University from 2007 to 2016. 
He visited the Energy Systems Division at Argonne National Laboratory, Argonne, IL, USA, as a visiting scholar from 2014 to 2015. 
He joined the Department of Electrical Engineering at Guangxi University in 2016, where he is an Associate Professor. 
His research interests include power system control, electric vehicles, and demand response.
\end{IEEEbiography}


\begin{IEEEbiography} [{\includegraphics[width=1in,height=1.25in,clip,keepaspectratio]{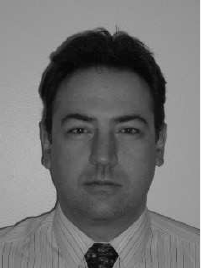}\vfill}]
{Aleksandar D. Dimitrovski} (SM)
received the B.Sc. and Ph.D. in electrical engineering with emphasis in power from the University Ss. Cyril \& Methodius, Macedonia, and M.Sc. in applied computer sciences from the University of Zagreb, Croatia. 

He is currently the Chief Technical Scientist in power and energy systems at the Oak Ridge National Laboratory, Oak Ridge, TN, USA, and also a Joint Faculty at the University of Tennessee, Knoxville. 
His research area of interest is focused on uncertain power systems, and their modeling, analysis, protection, and control.
\end{IEEEbiography}

\end{document}